\documentclass[lettersize,journal,compsoc]{IEEEtran}
\IEEEoverridecommandlockouts
\usepackage{makecell}
\usepackage{xcolor}
\usepackage{multirow}
\usepackage{multicol}
\usepackage{pifont}

\usepackage{amsmath,amsfonts}
\usepackage{algorithmic}
\usepackage{algorithm}
\usepackage{array}
\usepackage[caption=false,font=normalsize,labelfont=sf,textfont=sf]{subfig}
\usepackage{textcomp}
\usepackage{stfloats}
\usepackage{url}
\usepackage{verbatim}
\usepackage{graphicx}
\usepackage{cite}
\usepackage{tablefootnote} 
\renewcommand{\baselinestretch}{0.98}
\usepackage{xpatch}
\usepackage{xcolor}

\makeatletter

\def\changeBibColor#1{%
\in@{#1}{qin2018diagonalwise, onnxruntime}
\ifin@\color{black}\else\normalcolor\fi
}
\xpatchcmd\@bibitem
{\item}
{\changeBibColor{#1}\item}
{}{\fail}
\xpatchcmd\@lbibitem
{\item}
{\changeBibColor{#2}\item}
{}{\fail}
\makeatother
\begin{document}

\title{ReDas: A Lightweight Architecture for Supporting Fine-Grained Reshaping and Multiple Dataflows on Systolic Array}

\author{Meng~Han, Liang~Wang, Limin~Xiao, Tianhao~Cai, Zeyu~Wang, Xiangrong~Xu,  
Chenhao~Zhang
    \IEEEcompsocitemizethanks{
        \IEEEcompsocthanksitem Meng Han, Liang Wang, Limin Xiao, Tianhao Cai, Zeyu Wang, Xiangrong Xu and Chenhao Zhang are with the State Key Laboratory of Complex \& Critical Software Environment (CCSE) and School of Computer Science and Engineering, Beihang University, Beijing, China. Email:\{hanm, lwang20, xiaolm, caitianhao, wangzy1002, xxr0930, zch13021728086\}@buaa.edu.cn.
        \IEEEcompsocthanksitem *Liang Wang and Limin Xiao are the corresponding authors.
        
This work was supported in part by National Key R\&D Program of China under Grant 2023YFB4503100; in part by National Natural Science Foundation of China under Grant 62272026 and Grant 62104014; in part by Beihang Frontier Interdisciplinary Fund under Grant 
YWF-23-Q-1015, in part by the Academic Excellence Foundation of BUAA for PhD Students; and in part by the Iluvatar CoreX Semiconductor Company, Ltd. 
    }
}

\IEEEtitleabstractindextext{%
\begin{abstract}

The systolic accelerator is one of the premier architectural choices for DNN acceleration. However, the conventional systolic architecture suffers from low PE utilization due to the mismatch between the fixed array and diverse DNN workloads. Recent studies have proposed flexible systolic array architectures to adapt to DNN models. However, these designs support only coarse-grained reshaping or significantly increase hardware overhead. In this study, we propose ReDas, a flexible and lightweight systolic array that supports dynamic fine-grained reshaping and multiple dataflows. First, ReDas integrates lightweight and reconfigurable roundabout data paths, which achieve fine-grained reshaping using only short connections between adjacent PEs. Second, we redesign the PE microarchitecture and integrate a set of multi-mode data buffers around the array. The PE structure enables additional data bypassing and flexible data switching. Simultaneously, the multi-mode buffers facilitate fine-grained reallocation of on-chip memory resources, adapting to various dataflow requirements. ReDas can dynamically reconfigure to up to 129 different logical shapes and 3 dataflows for a $128 \times 128$ array. Finally, we propose an efficient mapper to generate appropriate configurations for each layer of DNN workloads. Compared to the conventional systolic array, ReDas can achieve about \textcolor{black}{4.6$\times$} speedup and \textcolor{black}{8.3$\times$} energy-delay product (EDP) reduction.

\end{abstract}

}

\maketitle

\IEEEdisplaynontitleabstractindextext

\IEEEpeerreviewmaketitle

\ifCLASSOPTIONcompsoc
\IEEEraisesectionheading{\section{Introduction}\label{sec:introduction}}
\else
\section{Introduction}
\label{sec:introduction}
\fi

Deep neural networks (DNNs) have demonstrated remarkable accuracy across a wide range of tasks. As AI-powered applications continue to advance, various fields such as autonomous driving, augmented reality (AR), virtual reality (VR), etc., are leveraging multiple deep neural networks to address diverse sub-tasks. The DNN model is constructed using a series of layers. Depending on the network topologies and structures of layers, different types of DNNs can be built such as Convolutional Neural Networks (CNNs), Recurrent Neural Networks (RNNs), Multilayer Perceptrons (MLPs), Transformers, etc. Additionally, several variants of layers have been proposed to enhance the DNN. For instance, variants of convolutional layers\cite{alzubaidi2021review} include 3D convolution, depth-wise convolution, deformable convolution, etc.

The significant heterogeneity in operations and shapes within and across DNN models leads to diverse workloads, posing a severe challenge in efficiently designing domain-specific accelerators. Due to their computation and memory-intensive nature, domain-specific accelerators have been actively developed to enhance DNN performance and energy efficiency\cite{TPUv2,Eyeriss,Simba,Diannao}. Systolic array architecture\cite{kung1979systolic} is one of the premier architectural choices for DNN acceleration. Benefiting from the regular 2D processing element (PE) array structure, systolic array inherently exploits the spatial data reuse and computation parallelism for general matrix multiplication (GEMM), which is the key operation of DNNs\cite{qin2020sigma}. However, fixed systolic arrays often exhibit low PE utilization under irregular workloads, indicating significant potential for architectural improvement. For many DNN layers such as Long Short-Term Memory (LSTM) and depth-wise convolution layer, the PE utilization can drop to less than 10\%\cite{lee2021dataflow}\cite{mapping}, keeping the vast majority of PEs idle. 

The fundamental cause of the low PE utilization comes from the mismatch between the fixed array and diverse DNN workloads. The shape (height and width of a 2D array) and dataflow are the two key properties of a systolic array. 
A promising solution should be capable of adapting to diverse DNN workloads by dynamically reconfiguring both the topological shape and the dataflow with a limited hardware overhead. We have observed that if the shape and the dataflow of an array can ideally adapt to DNN models layer by layer, it can achieve more than \textcolor{black}{6.3$\times$} speedup against a 128$\times$128 fixed architecture for EfficientNet-B0\cite{tan2019efficientnet} (detailed in Section~\ref{sec:diverse}). 

Recent studies have proposed flexible systolic array architectures to accommodate various DNN models. Unfortunately, prior works are still far from achieving this goal.
Gemmini\cite{gemmini} designs a flexible PE structure that supports both \textcolor{black}{Output Stationary (OS)} and \textcolor{black}{Weight Stationary (WS)} dataflows while the shape of the systolic array remains fixed. Planaria\cite{Planaria} enables coarse-grained reshaping of the systolic array to five different logical shapes under WS dataflow. Similarly, DyNNamic\cite{hanson2022dynnamic} can reshape the systolic array to various logical shapes under OS dataflow. However, both Planaria and DyNNamic only support a specific dataflow, which limits their efficiency in accommodating diverse DNN workloads. SARA\cite{SARA} is a recent work that allows for reconfiguration of both the systolic array shape and the dataflow. However, the dense dedicated links and multi-port buffer components of SARA incur significant overhead in terms of area and energy. Therefore, it is necessary to develop a flexible systolic array architecture that supports fine-grained reshaping and multiple dataflows, accommodating various DNN models while maintaining low design overhead.

In this paper, we present ReDas, a flexible and lightweight systolic array architecture that supports fine-grained reshaping and multiple dataflows. ReDas can dynamically reshape to up to 129 different logical shapes and 3 dataflows for a 128 $\times$ 128 PE array. To achieve this, we first propose reconfigurable roundabout data paths, which achieve fine-grained reshaping using only short connections between adjacent PEs and enable the data movement along two dimensions. Second, we design the microarchitecture of the PE and multi-mode data buffer. The PE structure introduces additional data bypassing and flexible data switching, which allow each PE to work at arbitrary dataflows and deal with the data from four directions. The multi-mode buffer supports fine-grained re-allocation of the on-chip memory resources to adapt to the requirement of different dataflows. Finally, to maximize the benefits of ReDas, we propose a mapping strategy that allows ReDas to adapt to various DNN models in a layer-by-layer manner. We have developed an elaborate analytical model to estimate the performance of ReDas under specific configurations. This analytical model takes into account additional constraints such as on-chip buffer capacity, off-chip bandwidth, as well as the ping-pong work mode. By considering these factors, we can more accurately identify the appropriate logical shape and dataflow for ReDas, ensuring efficient and effective utilization of the architecture.

We implement ReDas and evaluate it in eight DNN models. The results demonstrate that ReDas outperforms 
conventional systolic array with about \textcolor{black}{4.6$\times$} speedup and \textcolor{black}{8.3$\times$} energy-delay product (EDP), respectively. Compared to SARA, ReDas achieves about \textcolor{black}{2.1$\times$, 2.3$\times$} and \textcolor{black}{3.5$\times$} improvement in terms of power efficiency, energy-delay product (EDP) and area-delay-product (ADP), respectively.

To that end, the paper makes the following contributions:
\begin{itemize}
    \item We introduce a lightweight and reconfigurable roundabout data path using only the short connections between neighbor PEs. Compared to the dedicated bypass data path, the shared roundabout data path shows better scalability and lower overhead. 
    \item We present an efficient systolic architecture named ReDas by leveraging the reconfigurable roundabout data paths. By allowing the data movement along two dimensions, ReDas can flexibly support fine-grained reshaping and multiple dataflows.
    \item We propose ReDas Mapper to adapt to various DNN models. The mapper employs an elaborate analytical model and interval sampling to search the suitable hardware configuration and workload mapping.
    
\end{itemize}

The rest of the paper is organized as follows. In Section \ref{sec:2}, we introduce the necessary background of systolic array-based accelerators, related works and the core motivation of this work. In Section \ref{sec:architecture}, we describe the proposed architecture in detail. In Section \ref{sec:mapping}, we explain the mapping strategy for ReDas to adapt to various DNN models. In Section \ref{sec:result}, we present the experimental methodology and evaluation result. Finally, in Section \ref{sec:conclude}, we conclude this article.

\section{Background and Motivation} \label{sec:2}

\subsection{Heterogeneous DNN workloads} 

Currently, the integration of multiple DNNs within AI systems is becoming increasingly prevalent. DNNs are constructed using a series of layers. Each type of layer comprises a set of parameters and contains multiple variants, resulting in highly heterogeneous workloads in AI systems. For instance, Baidu Apollo autonomous system employs multiple DNN models in the sub-tasks of perception and prediction stages to enhance performance, such as traffic light detection, line detection, semantic segmentation and trajectory prediction\cite{raju2019performance}. The DNN workloads of the system involve various operations such as CONV2D, LSTM, MLP, \textcolor{black}{Fully Connected (FC)}, and \textcolor{black}{Depth-wise Convolution (DWCONV)}.

\textcolor{black}{While there are various DNN layers, general matrix multiplication (GEMM) remains the preferred abstraction. Most DNN layers can be transformed into GEMMs. In CNN, the convolutional layer can be transformed into a GEMM by the \textit{im2col} algorithm\cite{chellapilla2006high}, which unrolls high dimensional tensors into matrices. The FC layer is a GEMM without any transformation. In RNN, the LSTM layer contains 8 matrix-vector multiplications. The matrix-vector multiplication is a special case of GEMM. In the Transformer, the Multi-Head Attention (MHA) layer involves different numbers of GEMMs, depending on the number of heads of an MHA. The layer transformation process can be done during the DNN compilation phase to simplify hardware design to GEMM acceleration. Besides those linear layers, the DNN models involve non-linear layers (e.g., \textcolor{black}{Rectified Linear Unit (ReLU)}, softmax, sigmoid, tanh), which can not be transformed into GEMM, and they are treated as vector operations\cite{yang2021flexacc}.}

\textcolor{black}{Prior works\cite{xu2023survey, SystolicTensorArray, qin2020sigma} have shown that the GEMM operation is the performance bottleneck operation for DNN workloads. GEMM operations account for approximately 70\% of the total runtime during training\cite{qin2020sigma} and even higher during inference.} The heterogeneous nature of DNN workloads introduces both regular and irregular GEMM operations\cite{qin2020sigma}. For example, in Resnet-50\cite{he2016deep}, the dimensions (M, N, K) of GEMM operations can vary widely, with a total of 21 different combinations. These dimensions can range from (49, 2048, 512) to (12544, 147, 64). As a result, efficiently supporting heterogeneous GEMM workloads has become a crucial consideration in the design of DNN accelerators.

\begin{figure}[t]

    \centering
    \includegraphics[width=\linewidth]{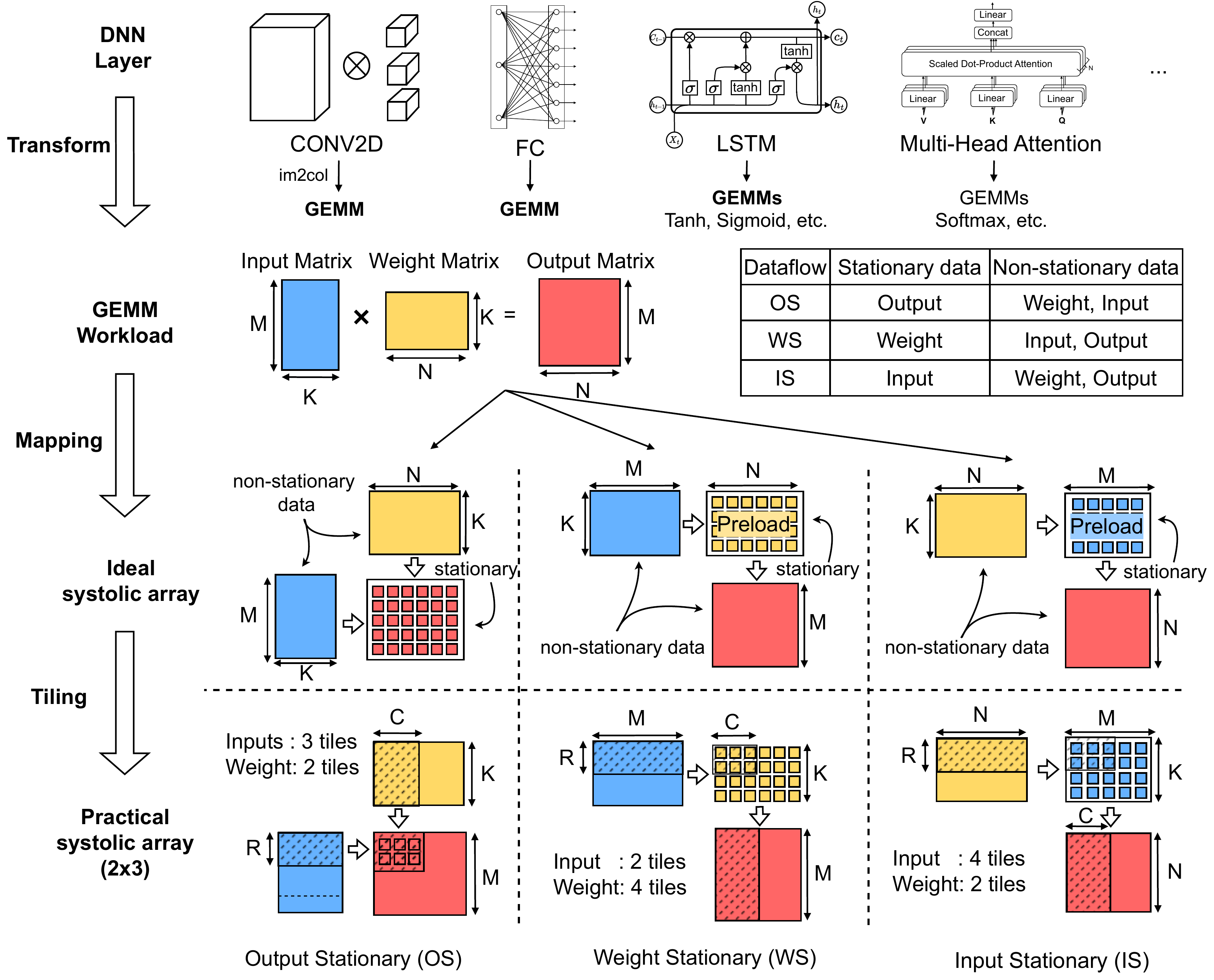}
    \caption{\textcolor{black}{Illustration of the execution process in the systolic array with different dataflows.}}
    \label{fig:illustration}
    \vspace{-0.4cm}
\end{figure}

\subsection{Systolic Arrays} \label{sec:background}

Systolic array is a premier architectural choice for accelerating DNN workloads. Systolic array architectures consist of two-dimensional arrays of PEs interconnected by peer-to-peer links. These architectures effectively exploit spatial data reuse and computation parallelism by enabling data transfer between neighbor PEs within the same rows and columns. Figure \ref{fig:illustration} provides an illustration of the conventional systolic array execution for DNN models. In this execution, a DNN layer such as CONV2D, FC, LSTM, and MLP is transformed into one or more GEMM operations. A GEMM operation involves multiplying an input matrix of size $M \times K$ with a weight matrix of size $K \times N$, resulting in an output matrix of size $M \times N$.

Dataflow and shape are the two key properties of the systolic array. There are three major kinds of dataflows for systolic array called Output Stationary (OS), Weight Stationary (WS) and Input Stationary (IS)\cite{Eyeriss}. Each dataflow maps one matrix (named stationary data) to the PE array while the remaining two matrices (named non-stationary data) are transferred cycle by cycle through the PE array in horizontal and vertical directions, respectively. The shape of a systolic array is another major property for GEMM execution. Designing a large enough systolic array which can map all the computing at once is costly. Practically, when the matrix size exceeds the practical systolic array shape, the matrix is divided into multiple tiles. Each tile will then be mapped onto the array sequentially until all outputs are generated. Figure \ref{fig:illustration} gives an example that the input matrix and weight matrix are split into multiple tiles under OS, WS, IS dataflows for a $2 \times 3$ systolic array. 

\subsection{Diverse Dataflow and Shape Requirements}\label{sec:diverse}

While the systolic array architecture inherently exploits spatial data reuse and computation parallelism for DNN workloads, it still faces challenges in efficiently handling diverse DNNs which result in low PE utilization\cite{mapping}. This issue is caused by the significant variation in computational characteristics, such as channel size and filter size, across different networks or even layers within single DNN model. Consequently, there is no one-size-fits-all configuration for a systolic array that achieves the best performance across all DNN workloads.

\begin{figure}[t]
    \centering
    \includegraphics[width=0.93\linewidth]{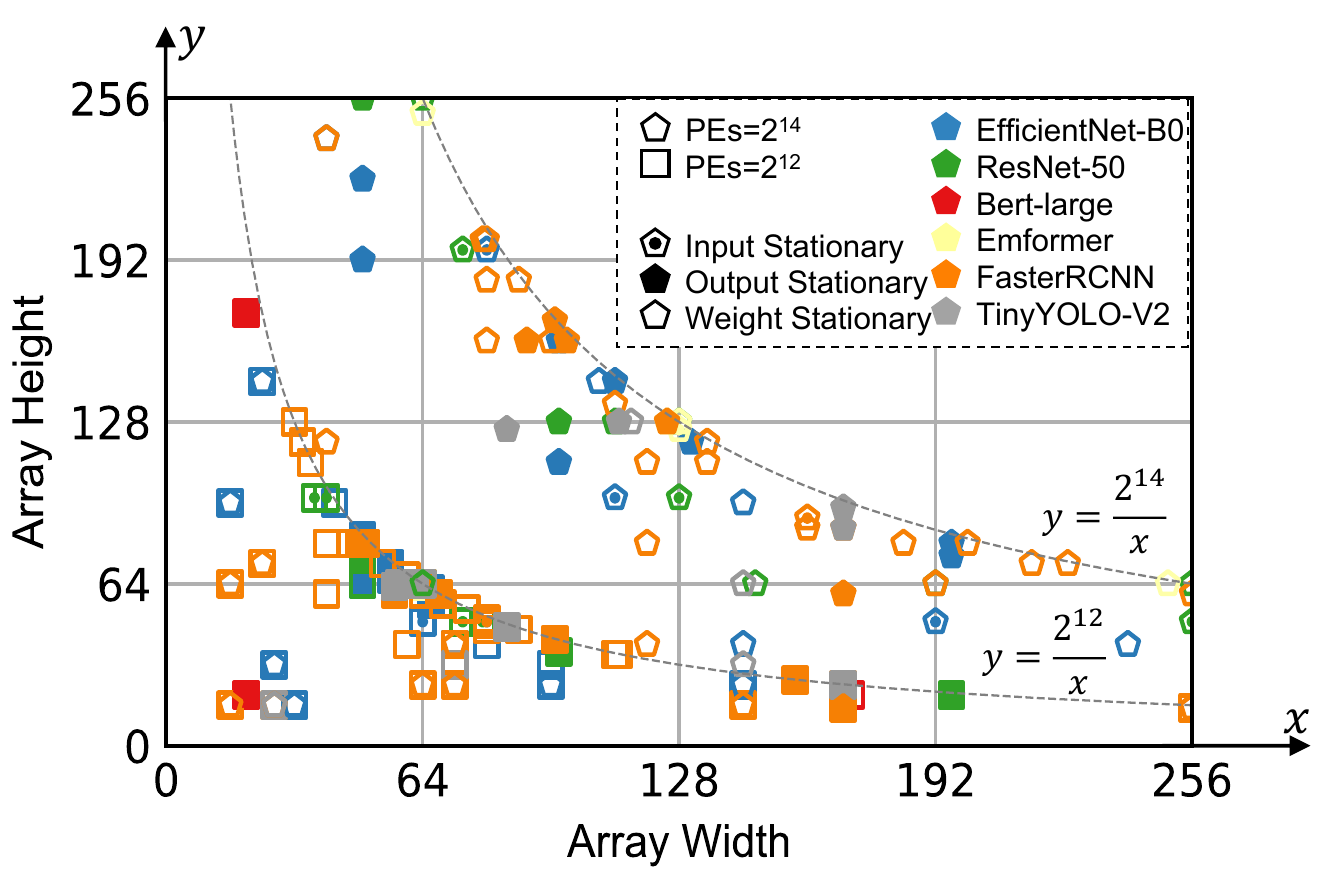}
    \caption{\textcolor{black}{Ideal dataflow and physical shape of systolic array vary with each layer of DNN models. Total number of PEs is not greater than $2^{12}$ or $2^{14}$.}}
    \label{fig:example}
\end{figure}

As shown in Figure \ref{fig:example}, we observe that the ideal\footnote{The "ideal" refers to the configuration that achieves the minimum execution time for the specific DNN layer.} array size (height and width of the physical shape) and dataflow (WS/OS/IS) vary from DNN layer to layer. To determine these ideal configurations, we explore all combinations of array shapes and dataflows within the constraints of the total number of PEs ($\leq$ $2^{12}$ or $2^{14}$) and available dataflows. The distribution of these configurations demonstrates a significant dispersion, highlighting the need for a flexible systolic array to support fine-grained reshaping and multiple dataflows.

\begin{figure}[t]
    \centering
    \includegraphics[width=0.93\linewidth]{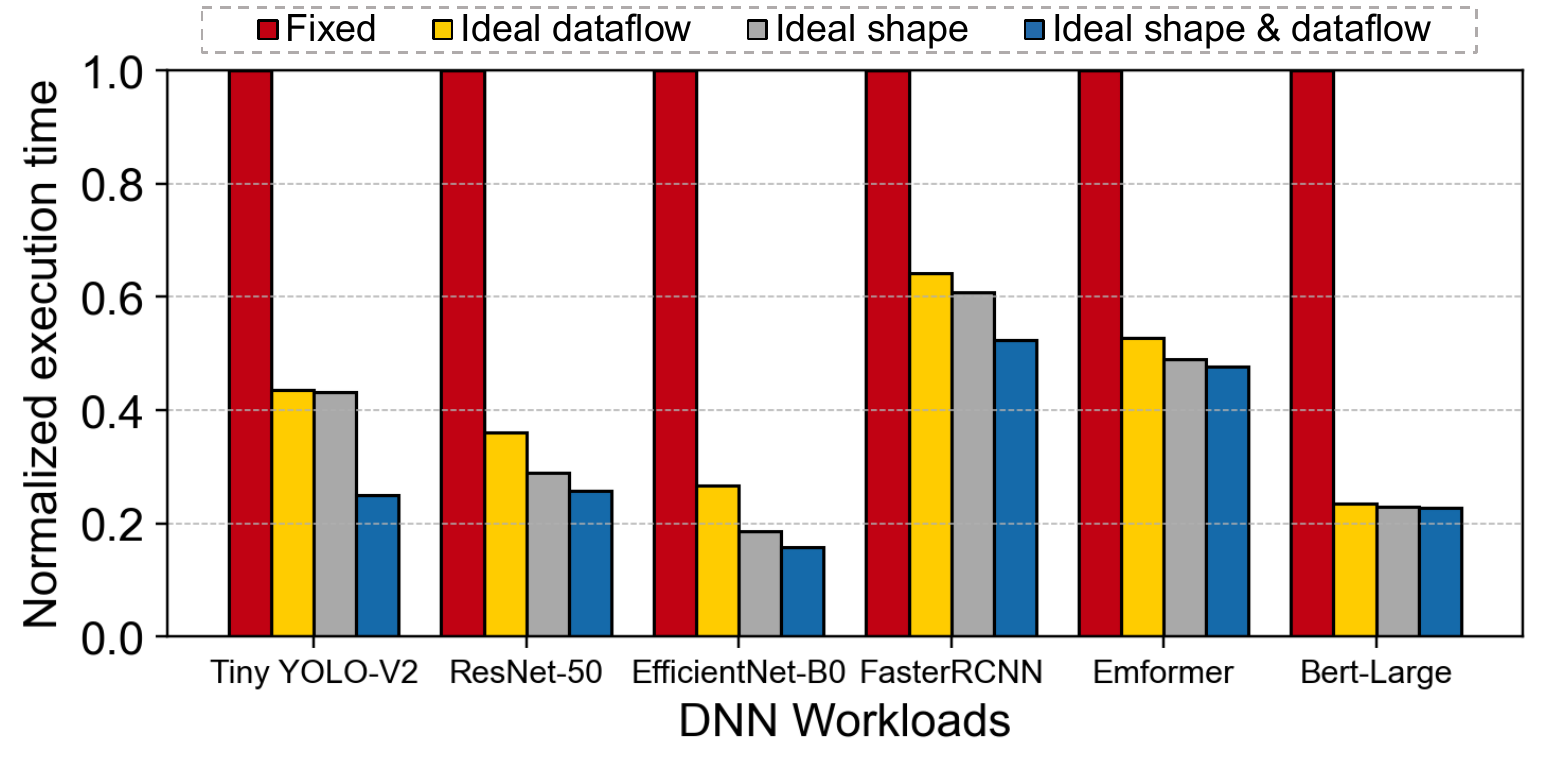}
    \vspace{-0.3cm}
    \caption{\textcolor{black}{Normalized execution time under different situation. \textbf{Fixed}: a 128$\times$128 fixed PE array with WS dataflow. \textbf{Ideal dataflow}: the dataflow (WS/OS/IS) of PE array is assumed to optimally adapt to DNN models layer per layer, the shape is fixed as 128 $\times$ 128. \textbf{Ideal shape}: the shape of PE array is assumed to optimally adapt to DNN models layer per layer, and the total number of PE is not greater than $128 \times 128$. The dataflow is fixed as WS.  \textbf{Ideal shape \& dataflow}: the shape and dataflow of PE array are assumed to optimally adapt to DNN models layer per layer.}}
    \label{fig:idea}
    \vspace{-0.4cm}
\end{figure}

As shown in Figure \ref{fig:idea}, another case study is conducted to show the significant potential of the flexible systolic array. We evaluate the execution time for running various DNN models in four situations. The case study shows that if the shape and the dataflow of an array can be ideally adapt to DNN models layer by layer, it can achieve more than \textcolor{black}{6.3$\times$} speedup against a $128\times 128$ fixed architecture for EfficientNet-B0. Moreover, the improvement is reduced if only one latitude of reconfiguration is supported.

\begin{table}[t]
\setlength\tabcolsep{4pt}
    \centering
     \caption{The Comparison of Flexible Systolic Array based DNN Accelerators.}
    \begin{tabular}{l|c|c|c|c}
    \hline 
         & \thead{Multiple \\ Dataflows} & \thead{Fine-grained \\ Reshaping} & \thead{Low Wire\\ Overhead}  & \thead{Low Buffer\\ Overhead} \\ \hline
       Gemmini\cite{gemmini}  & \checkmark &            & \checkmark  & \checkmark \\ \hline
       Planaria\cite{Planaria} &            &            & \checkmark   & \checkmark \\ \hline
       DyNNamic\cite{hanson2022dynnamic} &            &\checkmark  &            &            \\ \hline
       SARA\cite{SARA}     & \checkmark & \checkmark &  &            \\ \hline 
       ReDas (ours) & \checkmark & \checkmark & \checkmark & \checkmark  \\  \hline
    \end{tabular}
    \label{tab:comparison}
\end{table}

\subsection{\textcolor{black}{Related Works}}

\textcolor{black}{The heterogeneous nature of DNN workloads has driven the development of more flexible architectures. Several studies have proposed novel architectures that allow for dynamic reconfiguration of systolic array's dataflow and logical shape. Table \ref{tab:comparison} provides an overview of these studies. Gemmini\cite{gemmini} and Planaria\cite{Planaria} support multiple dataflows or coarse-grained reshaping of the systolic array. On the other hand, DyNNamic\cite{hanson2022dynnamic} and SARA\cite{SARA} support fine-grained reshaping along with one or multiple dataflows.}

\textcolor{black}{Gemmini introduces a flexible PE structure that supports both OS and WS dataflows, while keeping the shape of the systolic array fixed. Planaria breaks up the systolic array into multiple 32 $\times$ 32 sub-arrays and designs bypassing data paths between these sub-arrays to enable reshaping. However, Planaria only supports coarse-grained reshaping with a limited set of 5 logical shapes (without partitioning).}

\textcolor{black}{DyNNamic divides the convolution operations into shared kernels convolution and weighted accumulation stages, and designs a deformable systolic array under OS dataflow. It vertically splits the systolic array into sub-arrays and introduces additional bypassing data paths for inter-sub-array communication. Similar to Planaria, DyNNamic supports fine-grained reshaping under specific dataflow, limiting its flexibility for diverse DNN workloads.}

\textcolor{black}{SARA is a recent work that offers enhanced flexibility in both the systolic array shape and dataflow. It divides the PE array into multiple $4 \times 4$ sub-array and allocates dedicated links from on-chip buffer to the edge of each sub-array in both directions. Specially, the buffers are multi-ported with each bypass link allocated to separate ports. While the dedicated bypass links and multi-ported buffers of SARA enhance flexibility, they introduce significant overhead in wire usage, energy consumption, and power cost, leading to considerable increases in both power and area requirements.}

\subsection{Overhead of Flexibility} \label{subsec:overhead}

\begin{figure}[t]
    \centering
    \includegraphics[width=0.9\linewidth]{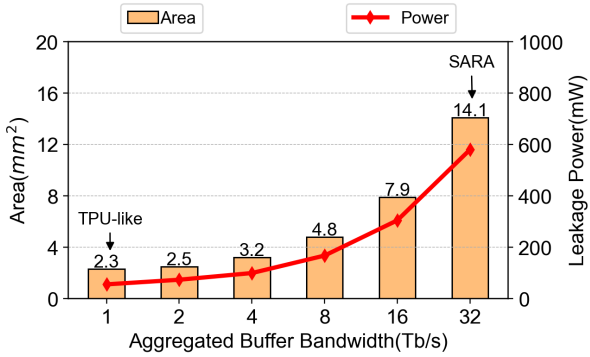}
    \vspace{-0.2cm}
    \caption{\textcolor{black}{The area and leakage power of 1MB multi-ported buffer under varying aggregated buffer bandwidth conditions. Assume the clock frequency is 1GHz.}}
    \label{fig:sramOverhead}
    \vspace{-0.4cm}
\end{figure}

Recent studies have explored diverse mechanisms to enhance flexibility in dataflow and reshaping capabilities for DNN accelerators. SARA and DyNNamic employed additional bypass links and multi-ported buffers to support fine-grained reshaping. However, these components significantly increase buffer and wire usage, thereby raising area and power consumption.

\textbf{Wire Cost:} Current designs often incorporate additional bypass links within the systolic array to enable dynamic reshaping of its logical shape. These bypass links play a crucial role in facilitating data movement among non-adjacent PEs and on-chip buffers. However, the interconnects pose significant challenges in the physical design process. We have conducted a place-and-route experiment for Gemmini-style and SARA-style PE array. The report reveals that the total wire length of a $32 \times 32$ SARA-style PE array is approximately \textcolor{black}{$9.5 \times$} longer compared to a Gemmini-style PE array. The bypass links are the predominant contributor to the extra wire length. The increased wire length results in higher capacitance and resistance, leading to increased delay in a quadratic trend\cite{lee2002optimal}. In addition, numerous BUFs are inserted within the interconnects to meet timing constraints\cite{chu1997closed}, causing additional area and power overhead.

\textbf{Buffer Cost:} SARA and DyNNamic utilize multi-ported buffers to address bandwidth limitations and optimize buffer capacity utilization. However, the multi-ported buffer comes with significant area and power costs. 

DyNNamic utilizes multi-ported SRAM to construct its multi-ported buffer. However, due to the modifications made to the SRAM bit cell, the area growth exhibits a quadratic relationship with the number of ports \cite{abdelhadi2014modular}. As a result, scaling up DyNNamic from a $20 \times 20$ size to a larger size, such as $100 \times 100$, would necessitate the use of 20-ports SRAM, posing practical challenges and limitations.

SARA provides dedicated SRAMs for each $4 \times 4$ sub-array in both directions. In particular, 1024 1KB SRAMs are employed to construct a 1MB 1024-ported buffer. Although such design is able to provide $32 \times$ bandwidth higher than TPU-like buffer, the overhead is significant. To quantify the area and power costs, we have conducted a case study following the buffer configuration in SARA. Several SRAMs with varying depths are employed to emulate a 1MB buffer under different aggregated bandwidth conditions. These SRAMs are generated by TSMC 28nm memory compiler. Figure \ref{fig:sramOverhead} shows that to meet the bandwidth requirement of SARA, the buffer area increases from \textcolor{black}{2.3} $mm^2$ to \textcolor{black}{14.1} $mm^2$, and the leakage power increases from \textcolor{black}{56} mW to \textcolor{black}{580} mW, significantly impacting the overall chip cost.

\subsection{Design Consideration}

The heterogeneous nature of DNN workloads demands a more flexible architecture. Numerous studies have introduced novel architectures enabling dynamic reconfiguration of systolic arrays in terms of dataflow and shape. However, these architectures often suffer from undesirable impact on area and power consumption.

To enhance efficiency and cost-effectiveness, exploring new systolic array architectures that support multiple dataflows and fine-grained reshaping is crucial. Such architecture should prioritize short connections among adjacent PEs to minimize the design cost. Additionally, it should take full advantage of the data reuse among PEs to reduce the overall bandwidth requirements for the on-chip buffer.

\section{Fine-grained Reshaping and Multiple Dataflows for Systolic Array} \label{sec:architecture}

\subsection{Architecture Overview}

\begin{figure}[t]
    \centering
    \includegraphics[width=0.75\linewidth]{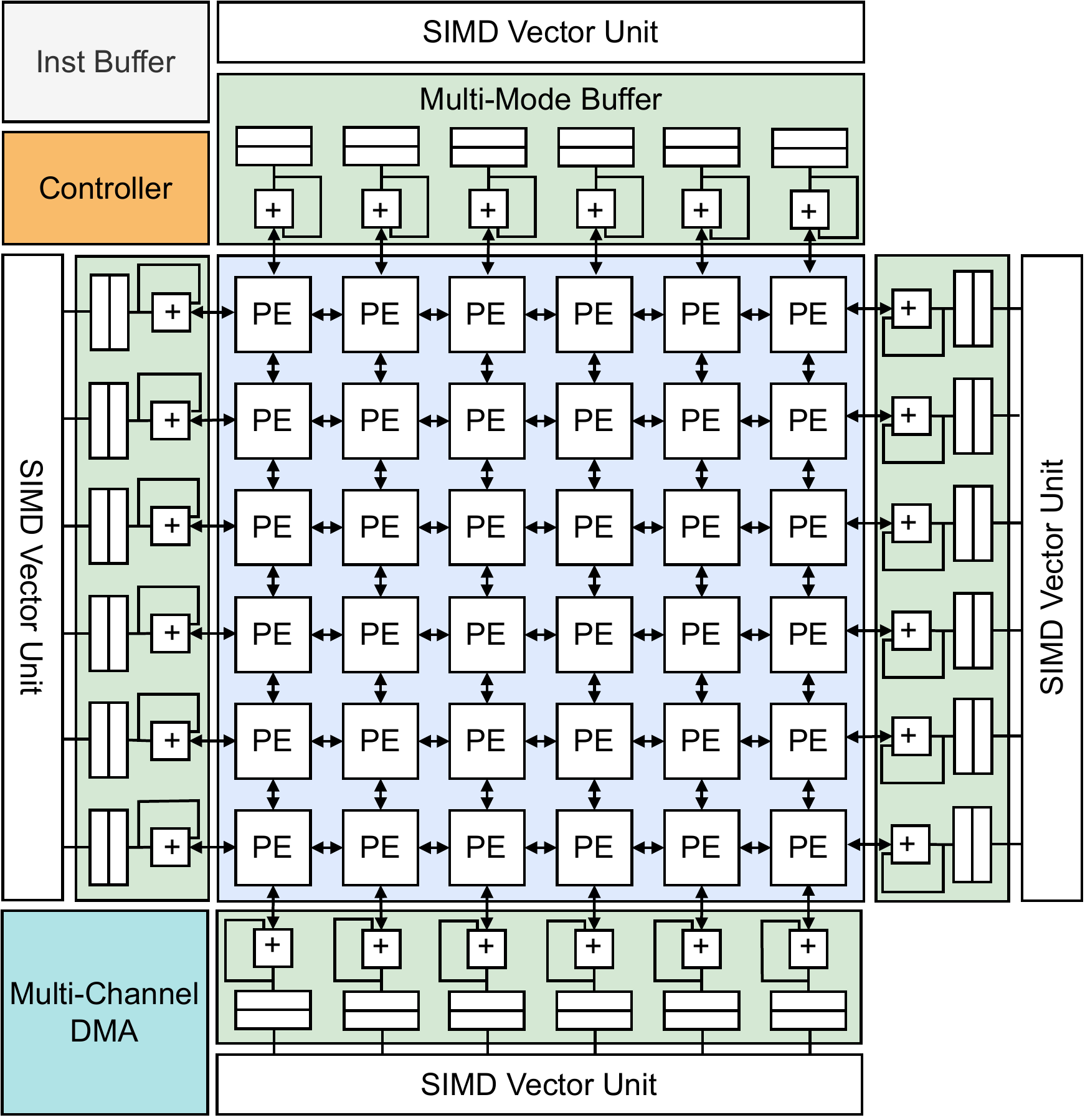}
    \caption{The overall architecture of ReDas.}
    \label{fig:overview}
    \vspace{-0.3cm}
\end{figure}

We present ReDas, a flexible and lightweight systolic array architecture that supports dynamic fine-grained reshaping and multiple dataflows. The overall architecture of the proposed accelerator is illustrated in Figure \ref{fig:overview}, which consists of four types of components, 1) a PE array, 2) four multi-mode buffers surrounding the PE array, 3) the SIMD vector units near the multi-mode buffers, and 4) other components, such as an instruction buffer, control unit, and DMA.

In the PE array, the neighboring PEs are interconnected using bidirectional links. These links enable the dynamic establishment of the roundabout data path, which is the key for supporting fine-grained reshaping of the systolic array. 

The multi-mode buffers connect to both the multi-channel off-chip DRAM and the PE array. Specifically, the multi-mode buffers consist of multiple banks, with each bank serving different roles depending on the dataflows and logical shapes being utilized. To support WS and IS dataflows, the accumulators are integrated in the multi-mode buffer. 

\textcolor{black}{ReDas employs four SIMD vector units to execute non-linear operations. In the runtime, the PE array and SIMD units work in a pipeline manner\cite{gemmini}. ReDas integrates NN-LUT\cite{yu2022nn} into SIMD units. NN-LUT is an accurate and hardware-friendly architecture that efficiently performs various operations (e.g., tanh, sigmoid, exponent, etc.) of the non-linear layers.
}

\subsection{Fine-Grained Reshaping with Roundabout Data Path}

\begin{figure}[t]
    \centering
    \includegraphics[width=\linewidth]{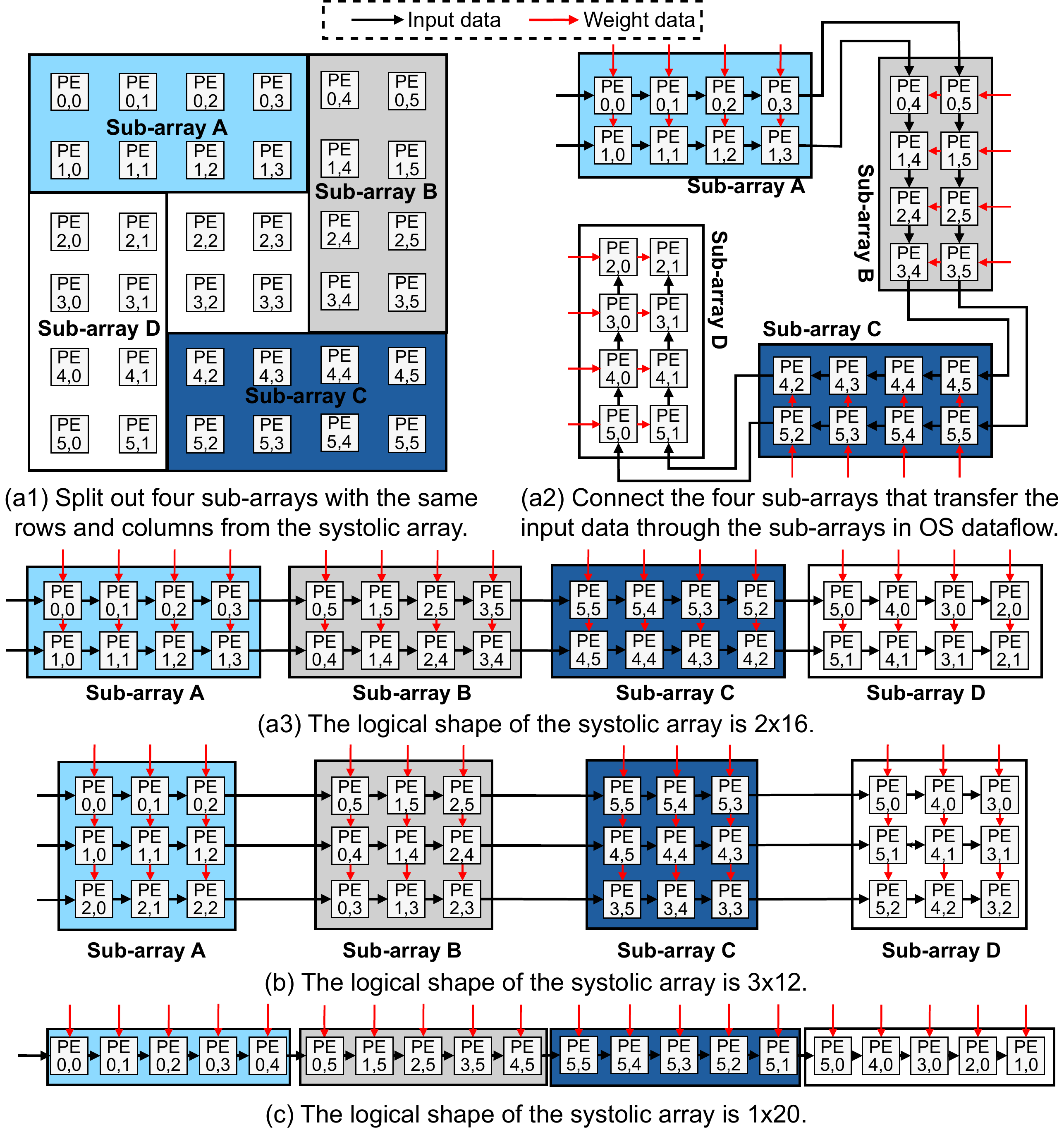}
    \caption{Illustration of ReDas execution on OS dataflow.}
    \label{fig:reshaping}
    \vspace{-0.4cm}
\end{figure}

The proposed systolic array can dynamically reshape into a number of logical shapes. 
Figure \ref{fig:reshaping} shows an example where a 6 × 6 systolic array is reconfigured into 3 of 7 possible logical shapes, which are $2\times 16$, $3\times 12$ and $1\times 20$ under OS dataflow. The key steps of reshaping into 2$\times$16 are shown in Figure \ref{fig:reshaping}(a1)(a2)(a3). In this case, the PE array is split into four sub-arrays, which are marked in different colors. And then, the sub-arrays are chained end-to-end using roundabout data paths, which enable the data movement  along 2 dimensions. In this manner, the input data, marked as the black lines, is transferred from the header PE in Sub-array A (PE[0, 0] and PE[1, 0]) to the tail PE in Sub-array D (PE[2, 0] and PE[2, 1]). The weight data is issued to the PE array from four edges, marked as red lines. Figure \ref{fig:reshaping}(a3) presents an equivalent logical shape with $2\times 16$ size. 
Other shapes can also be configured using the similar steps.

The roundabout data paths inherently support the fine-grained reshaping, as the sub-array can be allocated with various rows and columns of PEs. A logical shape is constructed by chaining 4 sub-arrays using the roundabout data paths.
Assuming the size of each sub-array is $R_s \times C_s$ with OS dataflow, the corresponding logical shape is $R_s \times 4C_s$ if the input data is transferred along the roundabout data path, or $4C_s \times R_s$ if the weight data is transferred along the roundabout data path. Figure \ref{fig:casesOfMultiMode} gives an example of reshaping to $2\times 16$ and $16  \times 2$ via the sub-array sized $2 \times 4$. 
Furthermore, to manage costs, the height of the sub-array should not exceed half of the physical PE array. Thus, a PE array with $R \times R$ size totally supports $R+1$ different logical shapes. 
For example, a $6\times 6$ array can be reshaped into 7 logical shapes of $1\times 20$, $20\times 1$, $2\times 16$, $16\times 2$, $3\times 12$, $12\times 3$ and $6\times 6$.

\begin{figure}[t]
    \centering
    \includegraphics[width=\linewidth]{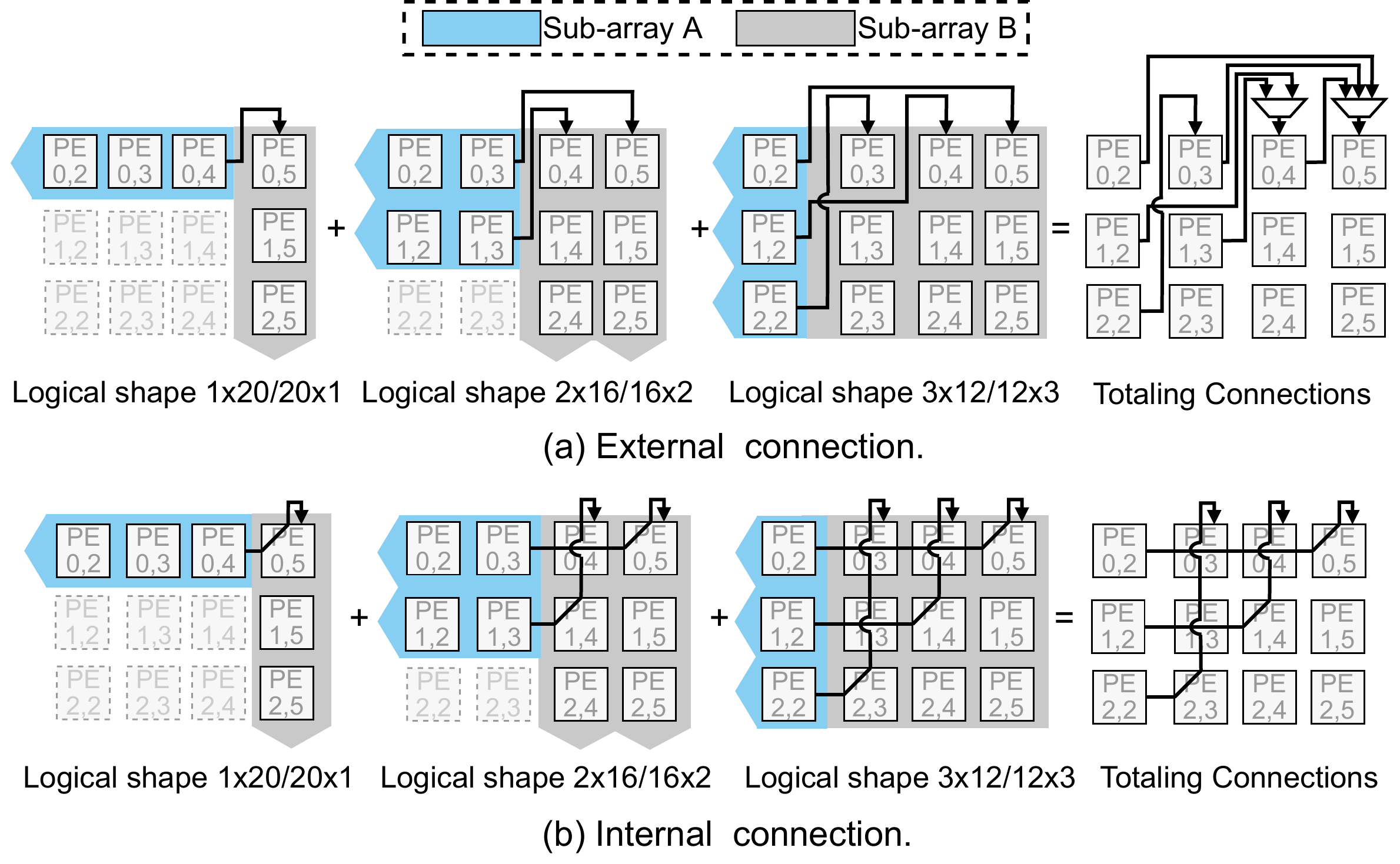}
    \caption{Two implementations of roundabout data path between the sub-array A and sub-array B in 6$\times$6 PE array. For better explanation, the figure only shows partial 6$\times$6 PE array.}
    \label{fig:roundabout}
\end{figure}

Figure \ref{fig:roundabout} shows two possible implementations of the roundabout data paths, i.e., the external connection manner and the internal connection manner. The external connection method directly connects the source and destination. However, this method is only suitable for a small-scale systolic array. Since there are multiple sources with the same destination, MUX units are employed to switch the sources. Moreover, the inevitable long connections break the regular short connections in the original systolic array.

ReDas uses the internal connection manner for the roundabout data paths as depicted in Figure \ref{fig:roundabout}(b). The internal manner can achieve a higher scalability and lower design cost. This is because the additional connections are only established among the adjacent PEs and are reusable under different logical shapes. The detailed design of the PE that supports internal connection manner is presented in Section~\ref{sec:PE}. Note that although the roundabout data paths may not use all the PEs in the systolic array, ReDas can still greatly improve the PE utilization up to 4 times in some cases compared to the conventional systolic array. \textcolor{black}{The detailed discuss is in Section \ref{subsec:sensitivity}.}

\subsection{Multiple Dataflows with Multi-mode Buffer} \label{sec:buffer}

\begin{figure}[t]
    \centering
    \includegraphics[width=0.95\linewidth]{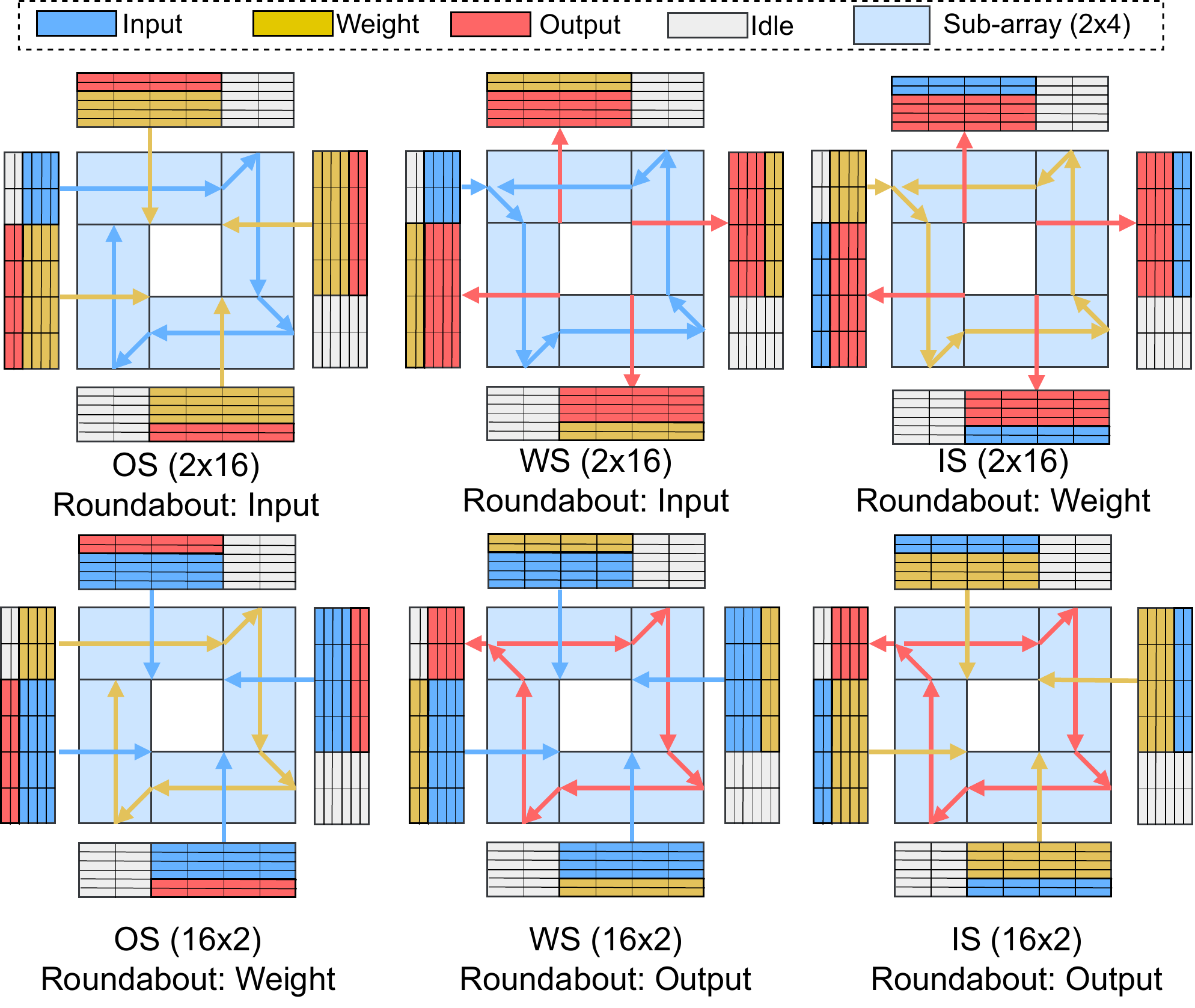}
    \caption{Examples of the on-chip buffers' working modes varying with the dataflow and logical shape. The dataflow, logical shape, and data in the roundabout data paths are represented in the diagrams. }
    \label{fig:casesOfMultiMode}
    \vspace{-0.4cm}
\end{figure}

ReDas efficiently supports multiple dataflows by leveraging the symmetrical patterns inherent in these dataflows. As discussed in Section \ref{sec:background}, for each dataflow, the PEs receive two non-stationary data and perform the \textcolor{black}{Multiplier-and-Accumulation (MAC)} operation with the stationary data scratched in the PE register. The PE can also flexibly adapt to three calculation patterns by adding an exchanging logic. 

Unlike the conventional systolic array architecture, ReDas employs four on-chip buffers called multi-mode buffers, which are positioned around the PE array. Each buffer consists of multiple independent banks. Such design easily supports dynamic fine-grained memory resource reallocation. In addition, to support WS and IS dataflow, the additional accumulators are integrated within each buffer (OS dataflow does not need accumulators).  

The type of operand stored in the multi-mode buffer is variable. In detail, both the stationary data and non-stationary data can be stored in the same buffer. During runtime, each bank within the buffer plays a special role of weight issuer, input issuer, output receiver or idleness depending on the specific dataflows and logical shape. Figure \ref{fig:casesOfMultiMode} illustrates the working modes of the buffers with different dataflows and logical sizes. The buffers are marked with four different colors representing the four working modes mentioned above. The solid arrows in the figure depict the movement of data. To optimize energy efficiency, when the buffer bank is set to an idle state, the SRAMs within the bank are switched to sleep mode for leakage power reduction. 

In the WS/IS dataflow, the PE array preloads the stationary data (weight/input data) from the multi-mode buffer in each direction. The data is transferred from the edges of the PE array towards the center. Once the data is loaded, the GEMM calculation begins. On the other hand, in the OS dataflow, at the end of the GEMM computation, the result data in each PE is transferred from the center to the edges of the PE array, and then it is stored back into the multi-mode buffers.

The fine-grained multi-mode buffer in ReDas offers two significant benefits. First, the multi-mode buffer allows for flexible allocation of operand data, enabling ReDas to meet the varying data requirements of different PE array logical shapes. For example, in the OS dataflow, when the PE array shape is $128 \times 128$, the input data and weight data have equivalent storage requirements. However, when the PE array is reshaped to $64 \times 256$, the input data requirement becomes four times that of the weight data. The multi-mode buffer can adapt to these changes and allocates the appropriate amount of buffer space, ensuring efficient utilization of resources. Second, the hybrid storage capability of the multi-mode buffer in ReDas improves the utilization of buffer bandwidth. This is achieved by storing both stationary data and non-stationary data within the same buffer. As the preloading of stationary data stage and the computation stage do not occur simultaneously. This ensures that the buffer remains busy during both stages, maximizing its utilization without the need for multi-port SRAM design.

\subsection{PE Structure} \label{sec:PE}

\begin{figure}[t]
    \centering
    \includegraphics[width=0.96\linewidth]{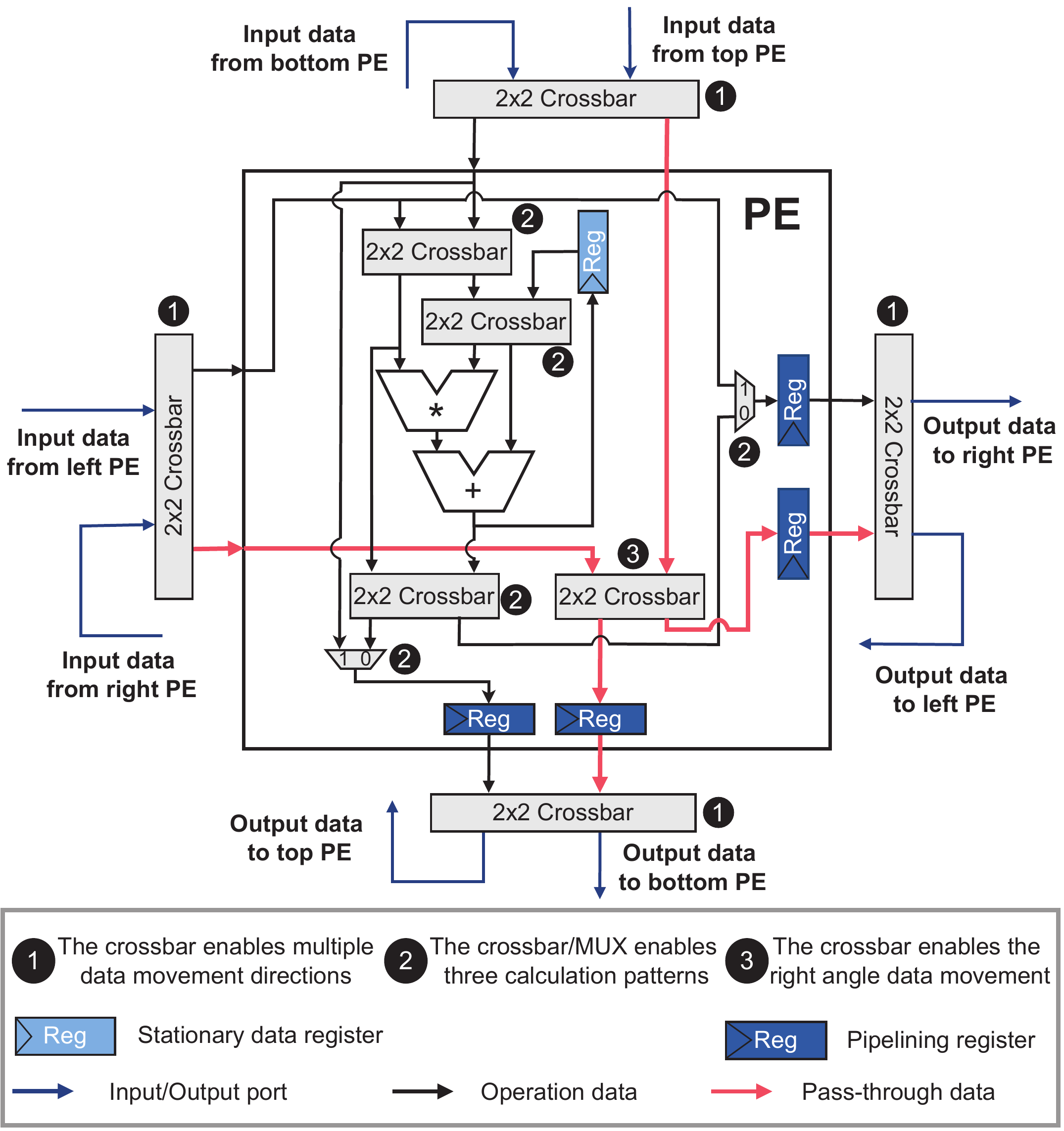}
    \caption{Structure of the PE.}
    \label{fig:PEstructure}
    \vspace{-0.4cm}
\end{figure}

The internal connection manner and three calculation patterns require a more flexible PE structure. Figure \ref{fig:PEstructure} shows the details of PE structure in our design. The MAC unit is used for the GEMM operation, as in the canonical PE. Furthermore, the ReDas PE features four input ports and four output ports, evenly distributed around the PE. The bidirectional connections between neighbor PEs are established to support the roundabout data path. 

The PE employs several crossbars and MUX units to support a variety of working modes. 
When a PE is employed to perform both the MAC operation and the roundabout data path (e.g., PE[0,4] and PE[1,3] of the logic shape $3\times 12$ in Figure \ref{fig:roundabout}(b)), the PE has up to four inputs simultaneously.
Two of them (named operation data) are the non-stationary data, which are sent to the MAC unit, and the other two (named pass-through data) are on the roundabout data path. The crossbars (\ding{182}) near the input ports differentiate the inputs into operation data and pass-through data. The pass-through data can either move straight or turn depending on the position of the current PE. 
The crossbar (\ding{184}) enables the selection of the 2 data movements. For the operation data, to satisfy the special input order of the three operands of the MAC unit, the crossbars (\ding{183}) above the MAC unit reorder these three operands. After the computation of MAC, the crossbar and MUX unit (\ding{183}) below the MAC unit recover these data to the original order. Finally, the crossbars (\ding{182}) near the output ports issue the pass-through data and operation data to destination ports.

\section{Configuration and Mapping} \label{sec:mapping} 
ReDas's high flexibility in dynamic reshaping and dataflow switching requires an efficient hardware configuration and GEMM mapping method during model compilation, crucial for its overall performance. We propose ReDas Mapper, a dedicated configuration and mapping engine to reach the best performance. \textcolor{black}{ReDas Mapper comprises three components: 1) a search space generator to create valid candidates of hardware configuration and GEMM mapping, 2) an analytical model for performance estimation, and 3) an interval sampling engine to efficiently reduce search space.} Table \ref{tab:terms} describes all the important terms used below.

\textcolor{black}{For a given DNN model, in the compilation phase, the model is transformed into a sequence of GEMMs. Then, for each GEMM workload, ReDas Mapper generates the search space. Since the size of the search space can be extremely large (exceeding $10^{10}$ for a single GEMM workload), ReDas Mapper employs interval sampling to prune the candidates. Finally, it estimates the runtime of each candidate using the analytical model and chooses the configuration and mapping with the minimal runtime. In the execution phase, ReDas sequentially runs every GEMM workload of the model. At the start of each GEMM workload, ReDas takes 128 cycles to configure the PE array and buffers according to the configuration information from ReDas Mapper.}

\begin{table}[t]
\setlength\tabcolsep{3pt}
\renewcommand{\arraystretch}{1.5}
    \centering
     \caption{Important Terms and Descriptions.}
    \begin{tabular}{c|l}
        \hline
            \thead{\textbf{Term}} & \thead{\textbf{Description}} \\ 
        \hline
            $R_p, C_p$& \makecell[l]{The number of physical PE array rows and columns.\tablefootnote{In this paper, we assume the physical PE array is square in shape, so that $R_p$ is equal to $C_p$.}} \\ 
            \hline
            $R_l$, $C_l$ & \makecell[l]{The number of logical PE array rows and columns.} \\ 
            \hline
            $Dataflow$ & \makecell[l]{The organization and movement of data within the \\ PE array, including WS, OS and IS.} \\
            \hline
            $D_{phy}$ & \makecell[l]{The \underline{phy}sical capacity of each multi-mode buffer bank.} \\ 
            \hline
            $D_{sta}, D_{non}$ & \makecell[l]{The allocated capacity of the multi-mode buffer bank for \\ the \underline{sta}tionary data tile and the \underline{non}stationary data tile .}  \\
            \hline
            $M$, $K$, $N$ & \makecell[l]{The GEMM workload dimensions. \\ Input activation matrix:  $M\times K$; \\ Weight matrix: $K\times N$; \\ Output matrix: $M\times N$.} \\ 
            \hline
            $M_t$, $K_t$, $N_t$ & \makecell[l]{The tile dimensions. \\ Input activation tile: $M_t \times K_t$; \\ Weight tile: $K_t \times N_t$; \\ Output tile: $M_t \times N_t$.} \\ 
            \hline
            $S_i, S_w, S_o$ & \makecell[l]{The data size of each type of tile. \\ Input tile size $S_i = M_t \times K_t$; \\ Weight tile size $S_w = K_t \times N_t$; \\ Output tile size $S_o = M_t \times N_t$.} \\
            \hline 
            $NUM_t$ & \makecell[l]{The number of iterations in which the PE array \\ computes the tiled GEMM workload. \\ $NUM_t = \lceil \frac{M}{M_t} \rceil \lceil \frac{K}{K_t} \rceil \lceil \frac{N}{N_t} \rceil$} \\ 
            \hline
            $T_r(s)$, $T_w(s)$ & \makecell[l]{DRAM access time for reading or writing data of size $s$.} \\
       \hline
    \end{tabular}
    \label{tab:terms}
\end{table}

\subsection{ReDas Search Space} \label{sec:mapspace}
\begin{figure}[t]
    \centering
    \includegraphics[width=0.75\linewidth]{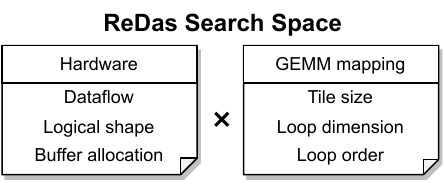}
    \vspace{-0.2cm}
    \caption{\textcolor{black}{The ReDas search space includes hardware configuration space and GEMM mapping space.}}
    \label{fig:searchSpace}
    \vspace{-0.4cm}
\end{figure}

\textcolor{black}{
As shown in Figure \ref{fig:searchSpace}, the ReDas search space is constructed from two orthogonal dimensions: 1) hardware configuration space, which includes logical array shape, dataflow, and buffer allocation; and 2) GEMM mapping space, encompassing tile size, loop dimension, and loop order. These dimensions significantly influence ReDas's runtime and power efficiency.}

\textbf{Logical shape and dataflow} determine the actual behavior of the PE array and how ReDas processes the operand matrices. Equation \eqref{eq_reshape_constraint} describes all the possible logical shapes for a ReDas array that supports PE-level reshaping granularity. 
\begin{footnotesize}
\begin{equation}
    \left\{ 
        \begin{array}{l}
            0 < R_l \leq \frac{1}{2}R_p \\
            C_l = 4 (C_p - R_l)
        \end{array}
    \right.
    \text{\textcolor{black}{or}} 
    \left\{
        \begin{array}{l}
            0 < C_l \leq \frac{1}{2}R_p \\
            R_l = 4 (R_p - C_l)
        \end{array}
    \right.
    \text{\textcolor{black}{or}} 
    \left\{
        \begin{array}{l}
            R_l = R_p \\
            C_l = C_p
        \end{array}
    \right.
    \label{eq_reshape_constraint}
\end{equation}
\end{footnotesize}

These logical shapes and three dataflows (WS, OS and IS) jointly construct the PE array configuration space, which is the key of ReDas search space that significantly influences the performance on diverse GEMM workloads.

\textbf{Buffer allocation} dictates how the multi-mode buffer bank is shared between the stationary operand and one of two nonstationary operands. The buffer allocation can significantly affect on-chip data reuse and potentially impact the overall runtime. For each bank, the total allocated capacity for these operands cannot exceed its physical capacity. The constraint is defined as
\begin{equation}
    \label{eq_buffer_sharing}
    D_{sta} + D_{non} \leq D_{phy}
\end{equation}

\textbf{Tile size} is the actual operand matrices' dimensions ($M_t$, $N_t$ and $K_t$) consumed by the PE array in each iteration. The valid tile size is constrained by the logical shape of ReDas array and the allocated buffer capacity. To maximize the PE utilization, ReDas Mapper sets two of the three dimensions (depending on the dataflow) equal to the logical array dimensions $R_l$ and $C_l$. The remaining dimension, however, has a direct influence on the first and last DRAM access overhead, which cannot be hidden by double buffering and becomes crucial on GEMM workloads that are not big enough to split into multiple tiles. It also has an impact on actual DRAM bandwidth utilization, which subsequently affects the runtime.

\textcolor{black}{\textbf{Loop dimension and order} determine the tile access and execution order of a GEMM workload. A GEMM workload is divided into a certain amount of tiles of identical size. The access and execution order of these tiles greatly impacts the on-chip buffer utilization and data reuse rate, subsequently affecting the duplicate DRAM accesses and the runtime.}

\textcolor{black}{Owing to its high flexibility and the range of GEMM mapping choices, the ReDas search space can become extremely large if generated naively without any reduction. For example, a $128 \times 128$ ReDas can support 129 logical shapes and 3 dataflows, and the physical capacity of a buffer bank is 4096 words. On a GEMM workload of dimensions (784, 256, 128), the number of valid candidates in search space is over $5.7 \times 10^{10}$. To improve the search efficiency of ReDas Mapper without losing much performance, we employ an interval sampling engine, which is discussed in Section \ref{sec:identifying}.}

\subsection{ReDas Analytical Model} \label{sec:analytical_model}
The ReDas analytical model estimates the overall runtime for a given GEMM workload at a specific hardware configuration and GEMM mapping. Equation \eqref{eq_double_buffering} describes the total runtime using a double buffering approach:
\begin{equation}
    \begin{aligned}
        T_{total} = T_{start} + NUM_t \times \max\{T_{exe}, T_{rd\&wt}\} + T_{end}
    \end{aligned}
    \label{eq_double_buffering}
\end{equation}
where $T_{start}$ is the time cost for reading the first input activation tile and weight tile from the off-chip DRAM and configuring the PE array, $T_{exe}$ is for the systolic array to compute one tile, $T_{rd\&wt}$ is for off-chip DRAM reading and writing for one set of tiles in the case that there is no on-chip reuse to avoid the DRAM access, $T_{end}$ is for writing the last output tile back to the off-chip DRAM. 

Equation \eqref{eq_t_systolic} represents the cycles cost for the systolic array to compute one tile with WS dataflow (each dataflow has a slightly different version of this equation).
\vspace{-0.1cm}
\begin{equation}
    \begin{aligned}
     T_{exe} = 
    \begin{cases}
        R_l + (R_l + C_l + M_t - 1) + 4 \times R_l & R_l < C_l \\
        C_l + (R_l + C_l + M_t - 1) + 4 \times C_l & R_l > C_l\\
        R_l + (R_l + C_l + M_t - 1) & R_l = C_l
    \end{cases}
    \end{aligned}
    \label{eq_t_systolic}
\end{equation}

The execution cycle is comprised of three parts. The first term represents the cycles for preloading a weight tile from the multi-mode buffer to each PE, with data being transferred from the edges of the PE array towards the center. The second term $(R_l + C_l + M_t - 1)$ is for consuming input activation tile and producing output tile. \textcolor{black}{The third term ($4\times R_l$ or $4\times C_l$) is the additional bypass cycles incurred by the roundabout data path when the logical shape differs from the physical shape. When $R_l < C_l$, ReDas requires $R_l$ cycles to rotate the flowing data by $90^\circ$ at each of the four corners, adding a total $4\times R_l$ to $T_{exe}$. When $R_l > C_l$, the additional bypass cycles amount to $4\times C_l$. In the case where $R_l = C_l$, the PE array operates as a conventional systolic array without reshaping, thus incurring no additional bypass costs.}

Equation \eqref{eq_t_dram} describes the DRAM access cycles $T_{start}$, $T_{end}$ and $T_{rd\&wt}$.
\begin{equation}
    \begin{aligned}
    \begin{cases}
        T_{start} = \max\{T_r(S_i) + T_r(S_w), R_p\} \\
        T_{end} = T_w(S_o) \\
        T_{rd\&wt} = T_r(S_i) + T_r(S_w) + T_w(S_o)
    \end{cases}
    \end{aligned}
    \label{eq_t_dram}
\end{equation}

\textcolor{black}{The value of $T_{start}$ is equal to the maximum of the data loading time and PE array configuration time.} At the beginning of a GEMM workload execution, ReDas loads the initial input activation and weight tiles from off-chip memory. Simultaneously, it performs the configuration of the PE array from top to down, which takes $R_p$ cycles. Because of the overlapping of these two operations, the hardware configuration barely causes extra cycles.

The analytical model takes on-chip tile reuse into consideration, which means the tiles already staged in the buffer do not need to be loaded again. This is implemented by a reuse-sensitive tile access sequence indicating whether each tile needs DRAM access, which is generated by ReDas Mapper before the runtime estimation.  

In practice, it is observed that the actual DRAM access efficiency largely depends on the data volume in a single DMA transaction, and hence the maximum DRAM bandwidth is too ideal for runtime estimation directly. An approximation method is proposed for more accurate estimation. We prerecord the actual DRAM access latency when reading and writing different amounts of data, and approximate the latency for accessing data of given size by linear interpolation, which is what function $T_{w}(s)$ and $T_{r}(s)$ implement in detail.

\subsection{Interval Sampling Engine}\label{sec:identifying}
ReDas Mapper employs an interval sampling engine to effectively reduce the search space for faster mapping while remaining near minimal GEMM workload runtime. 

ReDas Mapper utilizes interval sampling to aggressively lessen the buffer allocation and tile size choices. This reduction minimally impacts runtime, as operands are accessed and processed tile by tile. Additionally, ReDas Mapper avoids creating small tiles that would lead to significantly low PE utilization and DRAM access efficiency, further narrowing the search space. \textcolor{black}{Loop dimension and order significantly impact the buffer utilization. To prevent inefficient choices in loop dimension and order, ReDas Mapper generates loop nests based on the tile size and buffer allocation, so that the loop dimension and order are chosen carefully to maximize the buffer utilization.} Additionally, when a GEMM workload with the same dimensions appears, ReDas Mapper directly uses the previous choice, eliminating redundant search.  

\textcolor{black}{Benefiting from the interval sampling search engine, ReDas Mapper efficiently reduces the search space. For instance, in the case of ResNet-50, the size of the search space is reduced from 2.8$\times 10^{10}$ to an average of 1923 candidates per GEMM workload. On average, ReDas Mapper requires about 0.7 seconds per GEMM workload to estimate runtime and select the optimal configuration and mapping.}

\section{Evaluation} \label{sec:result}

\subsection{Experimental Setup}

\textbf{Benchmark} Following the MLPerf\cite{reddi2020mlperf} methodology, we choose a variety of DNN models from domains of image classification, object detection, machine translation and automatic speech recognition. The types of DNNs involved include CNN, RNN, and Transformer. The detailed characteristics of the benchmarks are shown in Table \ref{tab:benchmark}.


\begin{table}[t]
\setlength\tabcolsep{2pt}
\renewcommand{\arraystretch}{1.3}
    \centering
    \caption{Benchmarks.}
    \begin{tabular}{c|c|c|c|c}
    \hline
        \textbf{DNN Type} & \textbf{DNN Model} & \textbf{\# of Layer} & \makecell{ \textbf{Domain}}  & \textbf{Abbr.} \\ \hline
        \multirow{4}*{CNN }     & ResNet-50         & 54    & Image Classification  & RE \\ \cline{2-5}
                                & EfficientNet-B0   & 82    & Image Classification  & EF \\ \cline{2-5}
                                & TinyYOLO-V2       & 9     & Object Detection      & TY \\ \cline{2-5} 
                                & FasterRCNN & 46 & Object Detection & FR \\ \hline 
        \multirow{2}*{Transformer} & ViT            & 12    & Image Classification  & VI\\ \cline{2-5}
                                    & BERT-Large    & 24   & Machine Translation   &BE \\ \hline
                                    
        \multirow{2}*{RNN}          & GNMT          & 16    & Machine Translation   & GN\\ \cline{2-5}
                                    & DeepSpeech2   & 9    & \makecell{ Automatic Speech \\ Recognition}  & DS \\ \hline  
    \end{tabular}

    \label{tab:benchmark}
\end{table}

\begin{table}[t]
\renewcommand{\arraystretch}{1.3}
    \centering
    \caption{\textcolor{black}{ReDas configuration parameters.}}
    \begin{tabular}{l|c}
    \hline
        \makebox[0.1\textwidth][c]{\textbf{Parameter}} & \makebox[0.22\textwidth][c]{\textbf{Value}} \\ \hline 
        Systolic array size & 128x128 \\ \hline
        PE operating frequency & 700 MHz \\ \hline 
        PE data type & Int8 \\ \hline
        Technology & 28nm \\ \hline
        On-chip SRAM size & 4 MB \\ \hline
        Number of memory channels & 8 \\ \hline
        Memory bandwidth & 256 GB/S \\ \hline
        
    \end{tabular}
    
    \label{tab:baseline}
    \vspace{-0.2cm}
\end{table}

\textbf{Baseline} We use TPUv2, Gemmini, Planaria, DyNNamic, and SARA as the evaluation baselines. \textcolor{black}{We faithfully implement these architectures with Verilog HDL. The performance, area, power results are obtained by simulator and Synopsys tools.} In particular, TPUv2 supports WS dataflow. Gemmini supports both WS and OS dataflow while the systolic array shape does not support reshaping. Planaria supports the coarse-grained reshaping with WS dataflow. DyNNamic supports fine-grained reshaping with OS dataflow. SARA supports both fine-grained reshaping and multiple dataflows. 

\textbf{ReDas Implementation} ReDas accelerator is implemented with Verilog HDL and verified through RTL simulations. ReDas is synthesized with Synopsys tools in a 28nm process technology, with SRAMs generated by a memory compiler. \textcolor{black}{The ReDas Mapper is implemented in Python 3.8. It imports the DNN model's Open Neural Network Exchange (ONNX)\cite{onnxruntime} file and exports the ReDas hardware configuration and workload mapping strategy.} The same hardware parameters are used for the above baselines and ReDas for a fair comparison. The reshaping granularity of ReDas is limited to $4 \times 4$, which is consistent with SARA. The detailed parameters are shown in Table \ref{tab:baseline}. \textcolor{black}{We also implement the double-buffered on-chip memories for accelerators. These modifications highlight the impact of dataflow and reshaping capabilities rather than the low-level micro-architecture.}

\textcolor{black}{\textbf{Experimental Methodology} We closely follow the description of the original papers to reproduce the baseline hardware architectures and hardware configuration space. Because the detailed workload mapping strategies are unavailable for some baselines in original paper, we construct the GEMM mapping spaces and analytical models for accelerators and search for configurations with minimal runtime for a fair comparison. Moreover, for the depth-wise convolution layer, following existing implementation \cite{qin2018diagonalwise}, we rearrange multiple weight vectors into a large weight matrix, optimizing the PE utilization and performance of systolic array-based accelerators.} Following the methodology from prior work \cite{choi2020prema, lee2021dataflow, SARA, hanson2022dynnamic}, we evaluate performance in an extended SCALE-sim-v2 \cite{samajdar2020systematic}, a widely used cycle-accurate simulator for systolic array architecture. The simulator is verified against the Verilog implementation. We generate the Switching Activity Interchange Format (SAIF) files for every accelerator with real input data to estimate power consumption using the Design Compiler. DRAMsim3 \cite{DRAMsim3} is integrated with the simulator to model DRAM behaviors and energy consumption.

\vspace{-0.25cm}
\subsection{Performance Analysis}
\begin{figure}[t]
    \centering
     \includegraphics[width=\linewidth]{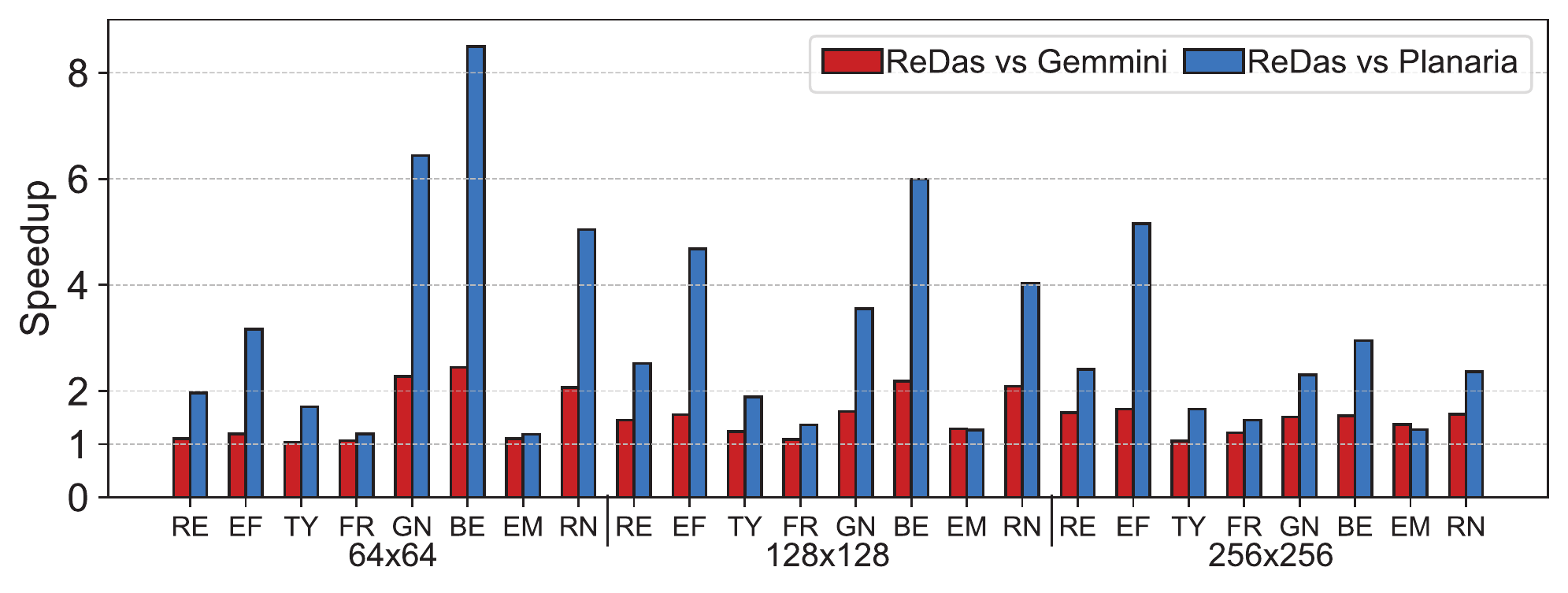}
                 \vspace{-0.3cm}
     \caption{\textcolor{black}{Performance of ReDas and other baseline accelerators on eight DNN workloads against the TPU.}}
    \label{fig:speedup}
\end{figure}

Figure \ref{fig:speedup} shows the normalized speedup brought by ReDas compared to baselines.  
ReDas achieves \textcolor{black}{4.6$\times$} speedup on average (geometric mean) against the TPU. For the baseline accelerators that only support multiple dataflow or coarse-grained reshaping, ReDas achieves \textcolor{black}{2.31$\times$} and \textcolor{black}{1.62$\times$} speedup on average, compared to Gemmini and Planaria. For the accelerators that support fine-grained reshaping, ReDas also achieves a \textcolor{black}{1.83$\times$} speedup over DyNNamic and comparable performance against SARA. 

Among all benchmarks, DeepSpeech2, GNMT, and ViT take the most significant benefit from our design, gaining \textcolor{black}{8.19$\times$, 5.66$\times$} and \textcolor{black}{6.01$\times$} speedup compared to TPU. On the one hand, the LSTM layers in DeepSpeech2 and GNMT, which tend to be transformed into GEMM operations with at least one dimension significantly smaller than that of the systolic array, are better suited to ReDas than to the baseline architectures. On the other hand, the feed-forward network (FFN) kernels are the major payload in ViT. They account for 55.1\% of the MAC operations. The GEMM dimensions of the FFNs, which are (50, 3072, 768) and (50, 768, 3072), enable ReDas to achieve a \textcolor{black}{7.5$\times$} speedup when reconfigured to a $52 \times 304$ logical shape under OS dataflow.

ReDas is slightly slower than SARA on GNMT workloads, with SARA achieving a speedup of \textcolor{black}{1.3$\times$} compared to ReDas. This is because of the numerous irregular but small GEMM workloads in these specific workloads. For instance, the matrix-vector multiplication in GNMT. SARA demonstrates higher performance in handling these workloads due to its shorter setup stage when operating under the parallel PE sub-arrays processing mode. However, such design requires a higher on-chip buffer bandwidth and additional dedicated links for every $4\times4$ PE array in each direction. As a result, SARA incurs higher power and area overheads. The detailed analysis is in Section \ref{subsec:ADP}.

\vspace{-0.2cm}
\subsection{Power Efficiency Analysis} 

\begin{figure}[t]
    \centering
     \includegraphics[width=\linewidth]{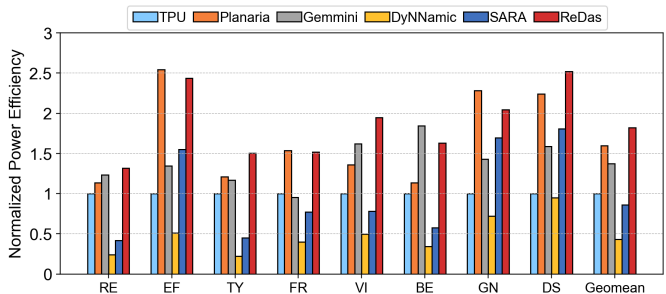}
              \vspace{-0.3cm}
     \caption{\textcolor{black}{Power efficiency of ReDas and other baseline accelerators on eight DNN workloads against the TPU.}}
    \label{fig:powerEfficiency}
    \vspace{-0.4cm}
\end{figure}

Figure \ref{fig:powerEfficiency} shows the power efficiency of accelerators on the evaluated workloads. All values are normalized to TPU. ReDas offers power efficiency improvements ranging from \textcolor{black}{1.32$\times$} to \textcolor{black}{2.52$\times$} compared to TPU. Compared to SARA, ReDas improves power efficiency, ranging from \textcolor{black}{1.2$\times$} to \textcolor{black}{3.3$\times$}, averaging \textcolor{black}{2.11$\times$}. This improvement is due to SARA and DyNNamic employing multi-ported buffers for fine-grained reshaping, significantly increasing SRAM read/write energy costs. In contrast, ReDas uses roundabout data paths and lightweight surrounding buffers, enhancing data reuse among PEs. As a result, ReDas achieves high power efficiency.

Among all benchmarks, Gemmini shows higher power efficiency than ReDas on BERT-Large workloads. In particular, Gemmini achieves a \textcolor{black}{1.13$\times$} improvement in power efficiency over ReDas. This is because the size of most GEMM operations in BERT-Large is larger than the size of PE array in both dimensions, such as (128, 1024, 4096), (128, 1024, 1024) and (128, 4096, 1024). Therefore, ReDas tends to operate under the 128 × 128 logical shape in OS/WS dataflow, similar to Gemmini's configuration, but the roundabout data paths in ReDas consume additional energy.

\vspace{-0.2cm}
\subsection{Area and Energy Breakdown} \label{subsec:area}

The ReDas is synthesized using Synopsys tools in a 28nm process technology, \textcolor{black}{specifically at the 0.81v@125C corner.} The synthesized report shows the critical path takes \textcolor{black}{0.87} ns, which means ReDas can run at up to \textcolor{black}{1.15} GHz.

\begin{table}[t]
\renewcommand{\arraystretch}{1.3}
    \centering
    \caption{\textcolor{black}{Area and Energy Breakdown.}}
    \begin{tabular}{l c c c c}
    \hline
        \makebox[0.1\textwidth][c]{\textbf{Module}} & 
        \makebox[0.04\textwidth][c]{\textbf{\makecell{ Area\\(mm$^2$)}}} & \makebox[0.04\textwidth][c]{\textbf{\makecell{ Area\\(\%)}}}  & 
        \makebox[0.04\textwidth][c]{\textbf{\makecell{ Energy\\(mJ)}}}  & 
        \makebox[0.04\textwidth][c]{\textbf{\makecell{ Energy\\(\%)}}}  \\ \hline 
        PE array    & 9.19     &   44.2    & 5.21 &    67.8\\
        ~~~~MACs    & 3.76      &   18.1    & 1.29 &    16.8\\
        ~~~~Original Muxes\&Regs\tablefootnote{The Muxes and registers (the stationary and non-stationary data registers) of PE in Gemmini-like systolic array.}    & 2.74  & 13.2  & 1.61 &    20.9 \\ 
        ~~~~Additional Muxes\&Regs  & 2.69  & 13.0   & 2.31 &    30.0  \\ 
        Multi-mode Buffers          & 10.21  & 49.2  & 1.05 &    13.7 \\ 
        ~~~~SRAM Macros             & 9.08  & 43.7  & 0.81 &    10.5 \\ 
        ~~~~Accumulators            & 0.37  & 1.8   & 0.03 &    0.4 \\ 
        ~~~~Bank Controllers        & 0.76  & 3.7   & 0.21 &    2.7 \\
        SIMD units                  & 0.85  & 4.1   & 0.18 &    2.3 \\
        Controller                  & 0.15  & 0.7   & $<0.01$ &   $<0.1$ \\
        Instruction buffer          & 0.06  & 0.3   & $<0.01$ &   $<0.1$ \\ 
        DMA                         & 0.31  & 1.5   & 0.22 &    2.9 \\ 
        Off-chip memory             & -     & -     & 1.01 &    13.1\\\hline
        Total                       & 20.77 & 100.0 & 7.69 & 100.0  \\ \hline
        
    \end{tabular}
    
    \label{tab:area_energy}
\end{table}

\begin{figure}[t]
    \centering
     \includegraphics[width=0.95\linewidth]{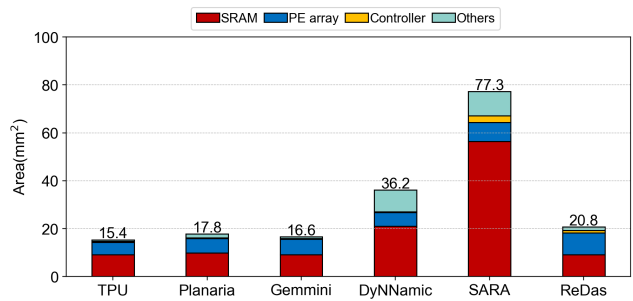}
         \vspace{-0.3cm}
     \caption{\textcolor{black}{On-chip Area for ReDas and other baseline accelerators.}}
    \label{fig:area}
    \vspace{-0.4cm}
\end{figure}

Table \ref{tab:area_energy} displays the synthesized area and energy breakdowns of ReDas for an inference on ResNet-50. Most of the accelerator footprint comes from the PE array and on-chip buffers. ReDas introduces additional components for fine-grained reshaping and multiple dataflows, such as mux units, registers, and bank controllers. Compared to the TPU architecture, these additional components increase area overhead by 35.3\%. Regarding power consumption, the synthesis report shows that the PE array in ReDas increases the power consumption by 149\% on average compared to the PE array in TPU. 
The power overhead mainly comes from the high PE utilization. Due to the ability of fine-grained reshaping and multiple dataflows, the PE utilization in ReDas is \textcolor{black}{4.75$\times$} higher than TPU. The more PEs involved in computation, the higher the power consumption is. As a result, ReDas achieves an improved power efficiency and energy efficiency. Compared to TPU, ReDas achieves about \textcolor{black}{1.8$\times$} power efficiency, \textcolor{black}{8.3$\times$} energy-delay product (EDP) reduction and \textcolor{black}{3.4$\times$} area-delay product (ADP) reduction.

\begin{figure}[t]
    \centering
     \includegraphics[width=0.9\linewidth]{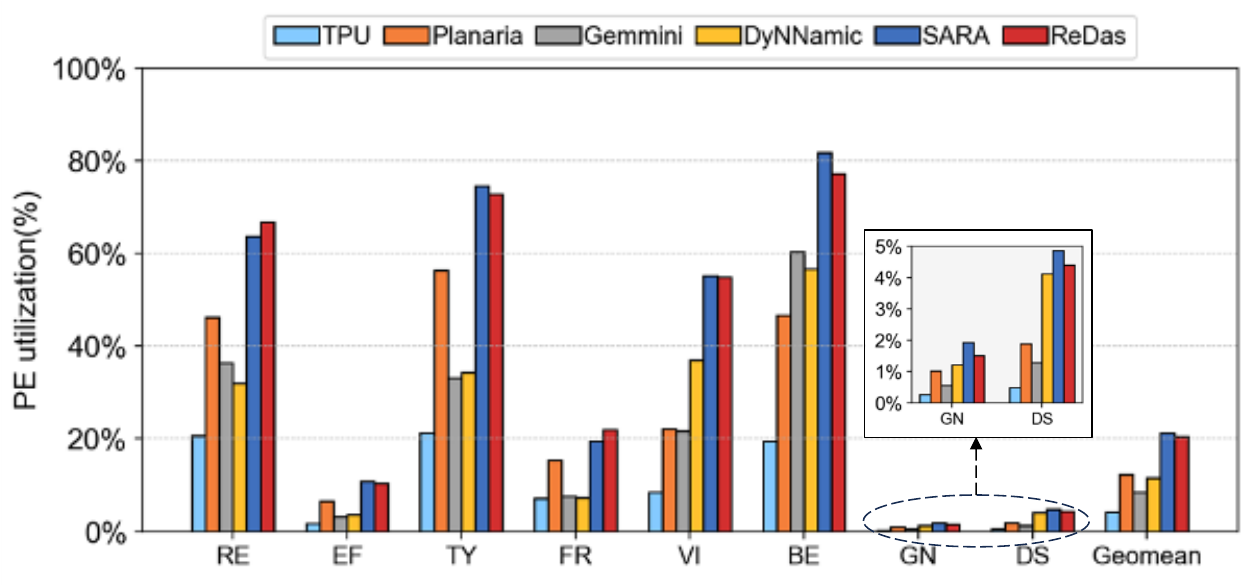}
     \vspace{-0.2cm}
     \caption{\textcolor{black}{PE utilization for different DNN models.}}
    \label{fig:spPEutilizationeedup}
            \vspace{-0.3cm}
\end{figure}

The distributed multi-mode buffers bring the insignificant overhead of die area and power. The synthesized report shows that the area of ReDas on-chip buffers is \textcolor{black}{10.21} $mm^2$. For comparison, the TPU-like concentrated buffers are \textcolor{black}{8.97} $mm^2$, and the SARA's multi-ported buffers are \textcolor{black}{56.47} $mm^2$. In the aspect of the buffer power consumption, ReDas distributed buffer is \textcolor{black}{4.19pJ/byte}, while the TPU-like concentrated buffer is \textcolor{black}{3.92pJ/byte}. However, the total energy consumption of SRAM access is roughly the same as TPU, as the roundabout data paths enhance the data reuse among the PEs, reducing the number of SRAM accesses. \textcolor{black}{In the aspect of off-chip memory, ReDas takes 13.31 pJ/byte on average to access data through HBM2.} 

Figure \ref{fig:area} provides an area breakdown comparison between ReDas and other baseline accelerators. The on-chip area is divided into four parts: SRAM, PE array, Controller, and Other. \textcolor{black}{The on-chip buffers and PE array are the predominant contributors to the area footprint. As for the computing components, the synthesis report shows that the PE array in ReDas increases the area by 77.4\% compared to the TPU-like PE array. The main reason is the additional muxes and registers within the PE array. Although the PE array raises the area overhead, the overall increase of the chip area is 35.3\% because the on-chip memory takes up a significant portion of the chip area.}   When compared to other accelerators such as DyNNamic and SARA, both of which also support fine-grained reshaping, ReDas demonstrates a much smaller overall area. For instance, ReDas takes up about \textcolor{black}{27\%} of the SARA area. The SRAM is the dominant contributor to the area footprint of SARA and DyNNamic.

\vspace{-0.2cm}
\subsection{PE Utilization Analysis}

We evaluate the PE utilization of the baseline accelerators and ReDas by running different DNN models. The PE utilization is defined as the ratio between the average number of activated PEs per cycle and the total number of PEs available in the architecture. The results are shown in Figure \ref{fig:spPEutilizationeedup}. ReDas can achieve \textcolor{black}{4.79$\times$, 1.67$\times$} and \textcolor{black}{2.42$\times$} higher PE utilization over TPU, Planaria and Gemmini on average, respectively. The PE utilizations for GNMT and DeepSpeech2 are significantly lower than for other DNN workloads, primarily because the major operations in these RNN models are matrix-vector multiplication operations. \textcolor{black}{Moreover, the PE utilizations for EfficientNet-B0 and FasterRCNN, which exploit depth-wise convolutions, are smaller than other CNN models. This is because the depth-wise 2-D filter is vectorized and mapped to a few columns of the array, even when optimized by filter gathering\cite{qin2018diagonalwise}.} 

\vspace{-0.2cm}
\subsection{Runtime Breakdown}
\begin{figure}[t]
    \centering
     \includegraphics[width=0.9\linewidth]{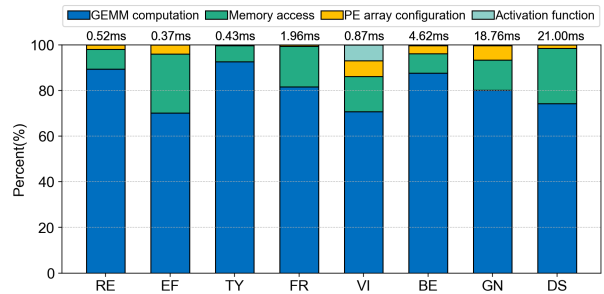}
     \vspace{-0.2cm}
     \caption{\textcolor{black}{Latency and runtime breakdown with different DNN workloads.}}
    \label{fig:runtimeBreakdown}
    \vspace{-0.3cm}
\end{figure}
\textcolor{black}{We further break down the runtime of DNN workloads into four parts, i.e., GEMM computation, Memory access, PE array configuration, and Activation function. As shown in Figure \ref{fig:runtimeBreakdown}, the execution of GEMMs significantly contributes to the total runtime. Due to the ping-pong work mode of the on-chip buffer, memory access operations are performed with GEMM operations simultaneously. As a result, only about \textcolor{black}{7\%} to \textcolor{black}{25\%} of the total runtime is non-overlapping. Furthermore, ReDas takes 128 cycles to configure the PE array and multi-mode buffers at the beginning of every GEMM execution, resulting in \textcolor{black}{0.4\%} to \textcolor{black}{7.0\%} runtime cost. The rest of the execution time is consumed by the non-linear activation and pooling layers, including max pooling, ReLU, sigmoid, softmax, and normalization, resulting in 0.1\% to 6.9\% runtime cost.} \textcolor{black}{We analyzed the cycles caused by the roundabout data path during the GEMM computations, as indicated by the additional bypass cycles of $T_{exe}$ in Equation (\ref{eq_t_systolic}). The results demonstrate that ReDas accounts for an average of 1.2\% of the total runtime, indicating minimal impact on performance.}

\begin{figure*}[t]
    \centering
     \includegraphics[width=0.95\linewidth]{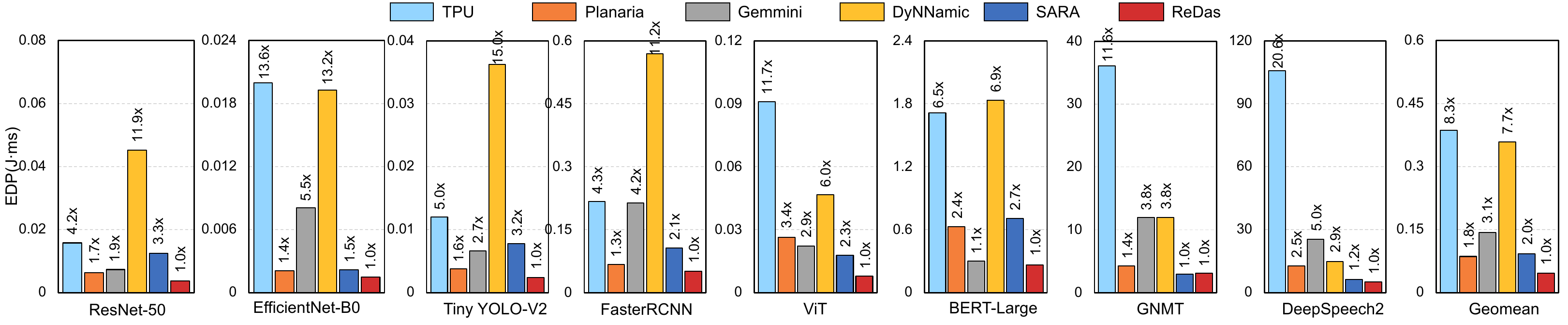}
     \vspace{-0.2cm}
     \caption{\textcolor{black}{Energy-delay product (EDP) of accelerators with different DNN workloads.}}
    \label{fig:EDP}
\end{figure*}

\begin{figure*}[t]
    \centering
     \includegraphics[width=0.95\linewidth]{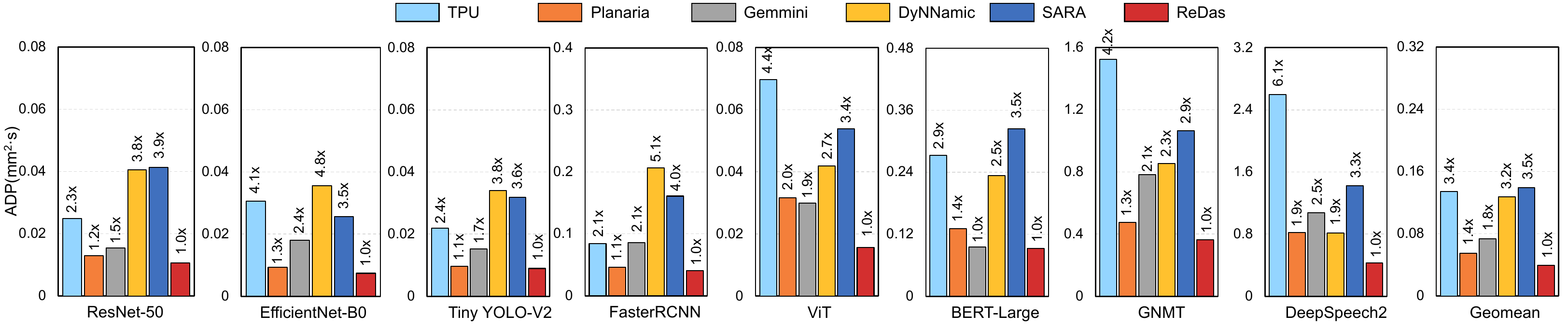}
     \vspace{-0.2cm}
     \caption{\textcolor{black}{Area-delay product (ADP) of accelerators with different DNN workloads.}}
    \label{fig:ADP}
    \vspace{-0.4cm}
\end{figure*}
\vspace{-0.2cm}
\subsection{EDP and ADP Analysis} \label{subsec:ADP}

To comprehensively evaluate the ReDas architecture, an analysis of the energy-delay product (EDP) and area-delay product (ADP) for each accelerator in the benchmark workloads is conducted. The EDP is calculated as the product of energy consumption and delay, while the ADP is the product of die area and delay. These metrics provide a comprehensive assessment of the trade-off between energy consumption, area cost, and computational efficiency. A lower EDP value indicates a more efficient balance between energy consumption and processing time, while a smaller ADP signifies a better balance between die area and delay.

In Figure \ref{fig:EDP}, the experimental results for the EDP are presented. The numbers on each colored bar in the figure are normalized to ReDas. On average, ReDas achieves \textcolor{black}{8.3$\times$} reduction in EDP compared to the TPU architecture. The lightweight design of ReDas contributes to a better balance of energy and performance when compared to SARA. Notably, ReDas exhibits EDP reduction up to \textcolor{black}{3.3$\times$} and \textcolor{black}{2.0$\times$ on average} compared to SARA across different benchmarks.

Figure \ref{fig:ADP} illustrates the ADP results. On average, ReDas achieves \textcolor{black}{3.4$\times$} reduction in ADP compared to TPU. As discussed in Section \ref{subsec:overhead}, previous designs aimed at supporting fine-grained reshaping and multiple dataflows often encountered high area overhead. However, ReDas successfully addresses this challenge, resulting in reduced ADP compared to DyNNamic and SARA architectures. On average, ReDas achieves an ADP that is \textcolor{black}{68\%} and \textcolor{black}{71\%} lower than DyNNamic and SARA, respectively. These results demonstrate the lightweight and effective nature of the ReDas architecture.

\subsection{Sensitivity Analysis}\label{subsec:sensitivity}

\begin{figure}[t]
    \centering
     \includegraphics[width=0.9\linewidth]{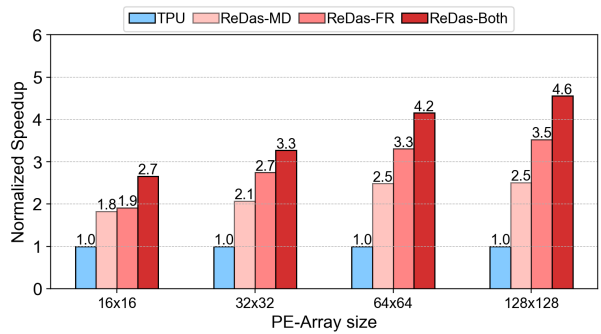}
        \vspace{-0.3cm}
    \caption{\textcolor{black}{Normalized performance of TPU and ReDas with different design points at several PE array scales. \textbf{ReDas-MD}: Only support dynamic reconfiguration for multiple dataflows. \textbf{ReDas-FR}: Only support fine-grained reshaping under WS dataflow.\textbf{ReDas-Both}: Support dynamic reconfiguration for multiple dataflows and fine-grained reshaping.}}
    \label{fig:sensiveAnalysis}
    \vspace{-0.4cm}
\end{figure}

We conducted sensitivity experiments to analyze the performance impact of the design points in ReDas architecture. 

\textbf{Hardware design points} Figure \ref{fig:sensiveAnalysis} illustrates the geometric mean speedup achieved by ReDas when supporting each or both design points, as compared to the TPU architecture. The experiment is conducted across four different PE array scales, ranging from $16\times16$ to $128\times128$.

Fine-grained reshaping and multi-dataflow reconfiguration contribute significant performance improvement, gaining \textcolor{black}{3.5$\times$} and \textcolor{black}{2.5$\times$} speedup with PE array size of 128$\times$128, respectively. Combining both design points pushes the enhancement even higher to \textcolor{black}{4.6$\times$} speedup, which demonstrates the effectiveness of ReDas's roundabout data path design and multi-dataflow support.

Furthermore, Figure \ref{fig:sensiveAnalysis} demonstrates a rising trend on performance improvement of ReDas compared to TPU as PE array size increases. This results from the fine-grained reshaping and multi-dataflow reconfiguration which render ReDas enough flexibility to utilize more PE for various DNN workloads when PE array grows larger. In contrast, TPU and other accelerators with low adaptability suffer from dramatic decrease of PE utilization.

\begin{figure}[t]
    \centering
     \includegraphics[width=0.9\linewidth]{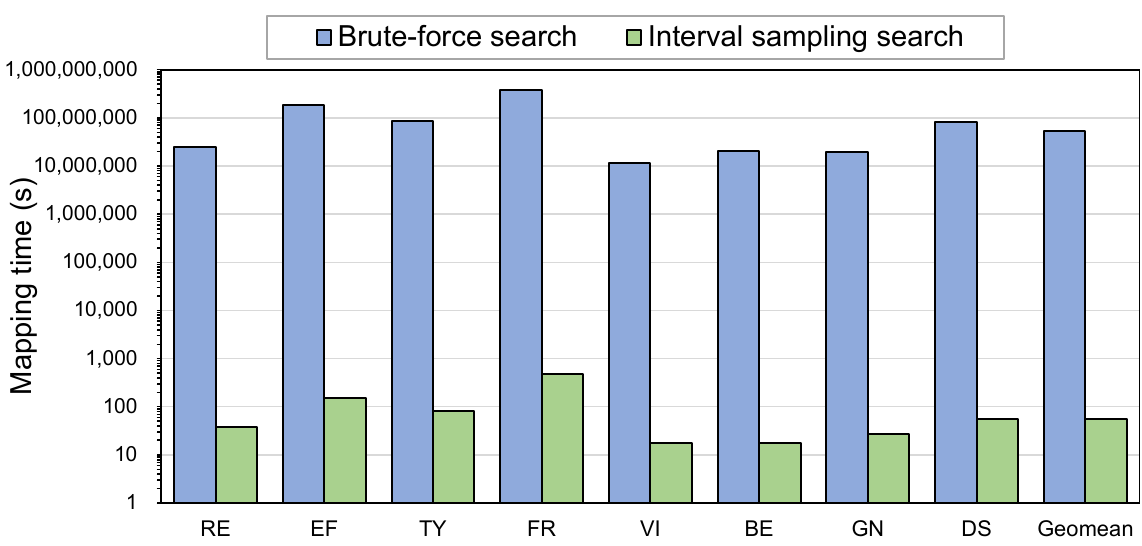}
    \caption{\textcolor{black}{The mapping time of brute-force searching and interval sampling searching with different DNN workloads\protect\footnotemark[4].}}
    \label{fig:mappingtime}
    \vspace{-0.4cm}
\end{figure}

\textcolor{black}{\textbf{Mapping process} Figure \ref{fig:mappingtime} illustrates the mapping time of brute-force searching and interval sampling searching with different DNN workloads. Due to the huge map space, the brute-force search takes several days to several months to generate the ReDas hardware configuration and workload mapping strategy. By contrast, on average, ReDas Mapper's interval sampling engine reduces the mapping time by six orders of magnitude. Regarding DNN workloads' execution time, the results of interval sampling search introduce a range of 0.1\% to 2\% performance loss compared to the results from brute-force search, showing the effectiveness of ReDas's interval sampling engine.} 

\begin{figure}[t]
\begin{minipage}[t]{0.35\linewidth}
    \centering
    \includegraphics[width=\linewidth]{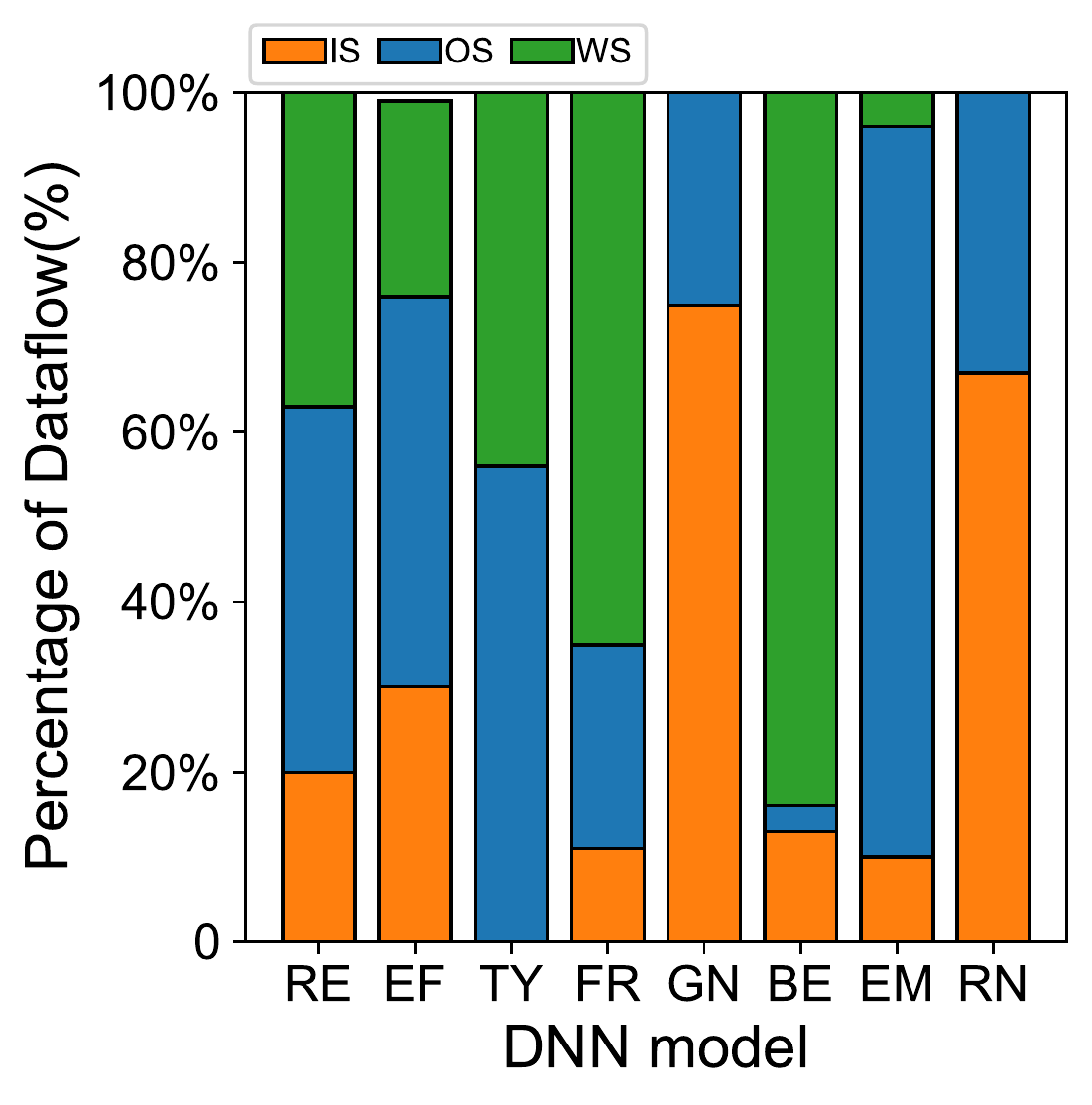}
    \caption{\textcolor{black}{Dataflow breakdown across DNN workloads.}}
    \label{fig:dataflows}
    \end{minipage}
    \hfill
    \begin{minipage}[t]{0.62\linewidth}
    \centering
    \includegraphics[width=\linewidth]{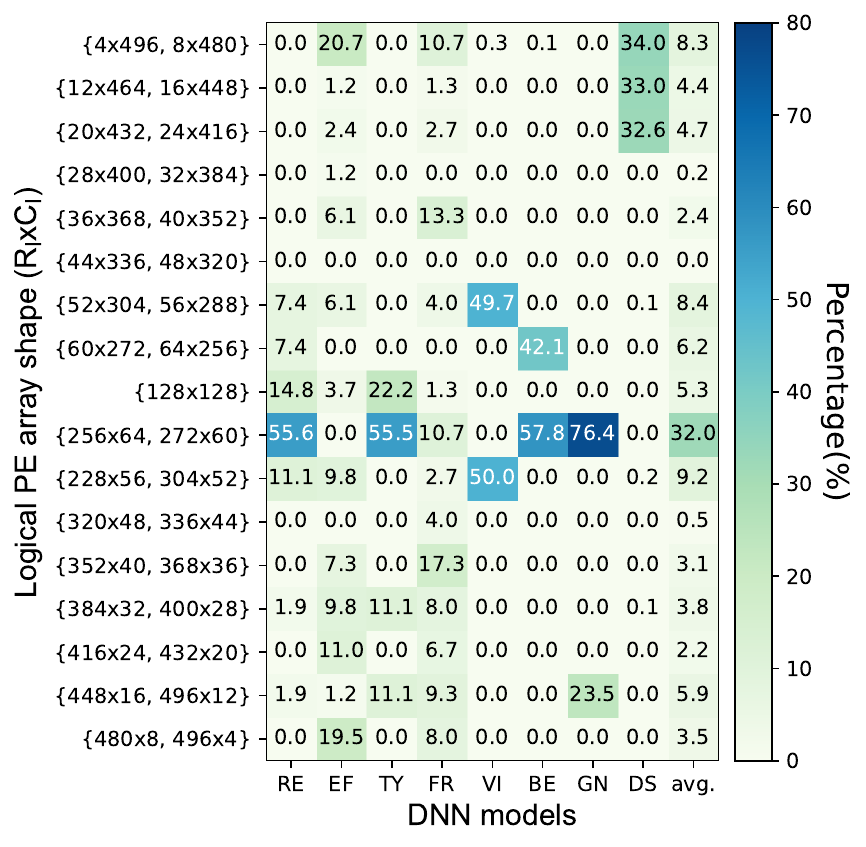}
    \caption{\textcolor{black}{Heatmap of ReDas's logical PE array shape across different DNN workloads.}}
    \label{fig:logicalShapes}
    \end{minipage}
    \vspace{-0.2cm}
\end{figure}

\textcolor{black}{\textbf{Dataflow and logical PE array shape} Figure \ref{fig:dataflows} and Figure \ref{fig:logicalShapes} illustrate the distribution of dataflows and logical shapes during ReDas executing various DNN workloads. ReDas operates approximately 40.9\% of DNN layers using OS dataflow and 39.7\% using WS dataflow. Concerning logical PE array shapes, the 256x64 configuration is the most prevalent, representing 27.3\% DNN layers. GNMT, BERT-Large, and ResNet-50 are the top three DNNs utilizing this configuration. Other logical shapes are also employed, ranging from 0.2\% to 9.2\% of DNN layers. These distributions highlight the importance of supporting fine-grained reshaping and multiple dataflows.} 

\begin{figure}[t]
    \centering
     \includegraphics[width=\linewidth]{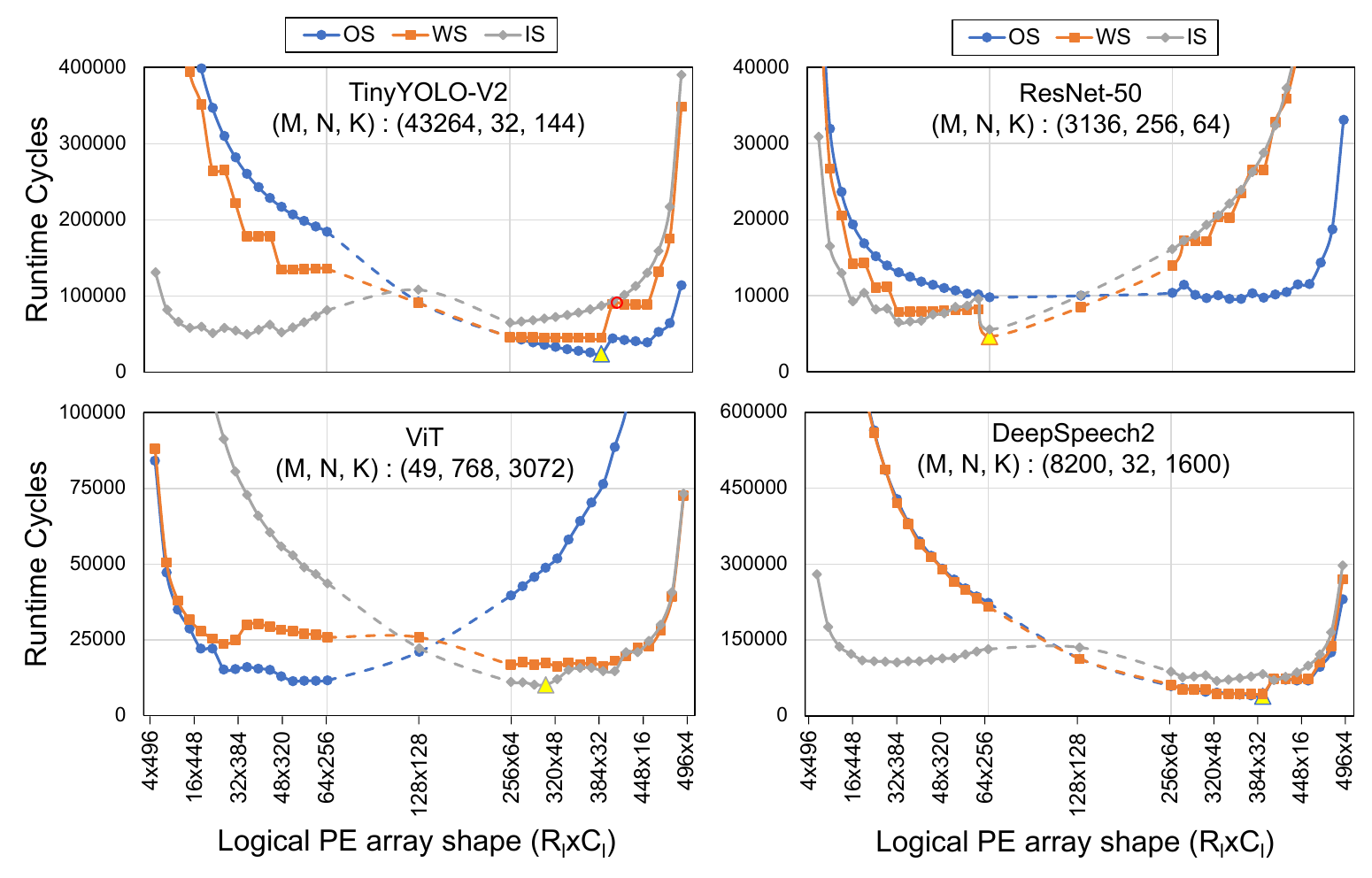}
    \caption{\textcolor{black}{The runtime of DNN layers with different logical shapes and dataflows.}}
    \label{fig:designPoints}
    \vspace{-0.4cm}
\end{figure}

\textcolor{black}{We conducted a case study to illustrate how ReDas achieves high performance through fine-grained reshaping and multiple dataflows. In Figure \ref{fig:designPoints}, the runtime of four DNN layers is shown with various logical shapes and dataflows. We utilized ReDas Mapper to generate the remaining configurations for a specific dataflow and logical shape. The yellow triangles highlight the dataflow and logical shape with the minimum runtime. Performance varies almost continuously with different logical shapes, and ReDas achieves optimal performance with various configurations across DNN layers. Although the roundabout data paths may not use all the PEs in the systolic array, ReDas can gain performance through fine-grained reshaping. For instance, as depicted in the top-left subfigure in Figure \ref{fig:designPoints}, the GEMM dimension of the second layer of TinyYOLO-V2 is (43264, 32, 144), ReDas achieves optimal performance when reshaping to 384$\times$32 with OS dataflow, where 75\% of PEs are involved in GEMM execution. In comparison, when mapping the workload to the conventional systolic array (128x128, OS), only the first 32 columns of the PE array (25\% of PEs) are involved in GEMM execution. As a result, the ReDas sized 384$\times$32 achieves 3.79$\times$ speedup compared to the ReDas sized 128$\times$128.}

\footnotetext[4]{\textcolor{black}{The mapping procedures are executed on the Intel Xeon Gold 5218R CPUs operating at 2.1 GHz. The mapping time of the brute-force search is the CPU time consumed by a multi-threaded program.}}

\vspace{-0.3cm}

\section{Conclusion}\label{sec:conclude}

We proposed ReDas, a flexible and lightweight systolic array to support fine-grained reshaping and multiple dataflows. Through the innovative design of the roundabout data path, the PE structure, and the multi-mode buffer, ReDas is capable of adapting to DNN layers by dynamically reconfiguring the logical shapes and dataflows of the systolic array, achieving this flexibility at a low cost. The evaluation results demonstrate that, in comparison to conventional systolic arrays, ReDas achieves about \textcolor{black}{4.6$\times$} speedup and \textcolor{black}{8.3$\times$} energy-delay product (EDP) reduction. \textcolor{black}{For future work, we aim to explore the performance benefits of accelerating structurally sparse neural networks\cite{mao2017exploring}\cite{he2017channel} 
and to investigate the application of advanced power management techniques, such as dynamic voltage and frequency scaling (DVFS), 
to further optimize the energy efficiency of the ReDas architecture in various scenarios. }

\bibliographystyle{IEEEtran}
\bibliography{references}

\begin{IEEEbiography}[{\includegraphics[width=1in,height=1.23in,clip,keepaspectratio]{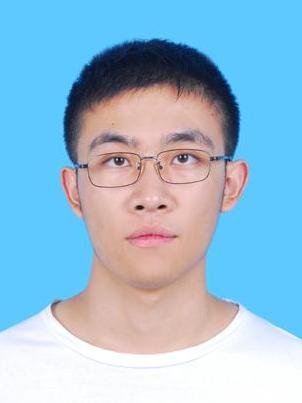}}]{Meng Han}
received the B.S. degree in Computer Science from the Beijing University of Posts and Telecommunications, Beijing, China, in 2019. He is currently working toward the Ph.D. degree in Computer Architecture with the School of Computer Science and Engineering, Beihang University, Beijing, China. His research interests include computer architecture, deep learning accelerator and high performance computing.
\end{IEEEbiography}
\vspace{-1.2cm}

\begin{IEEEbiography}[{\includegraphics[width=1in,height=1.2in,clip,keepaspectratio]{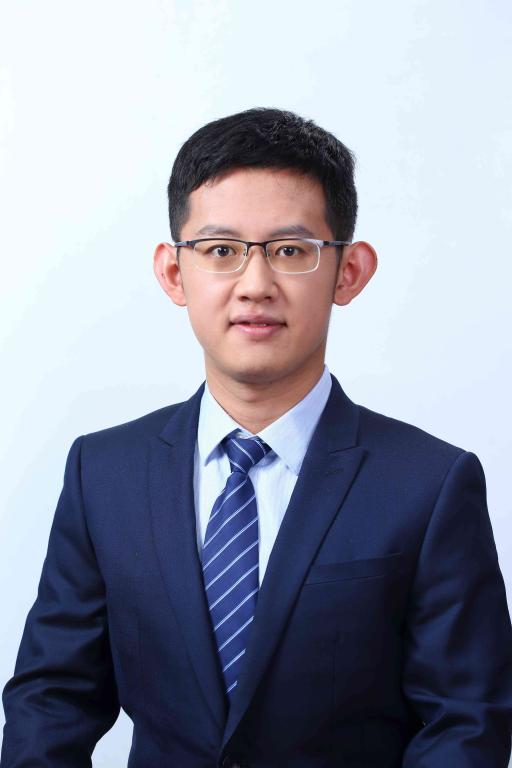}}]{Liang Wang} received the BEng and MSc degree in electronics engineering from Harbin Institute of Technology, China, in 2011 and 2013 respectively, and the Ph.D degree in Computer Science and Engineering from The Chinese University of Hong Kong in 2017. He is currently an assistant professor with the School of Computer Science and Engineering, Beihang University, China. He was a postdoctoral research fellow in Institute of Microelectronics, Tsinghua University from 2018 to 2020. His research interests include computer architecture, heterogeneous computing systems, etc. \end{IEEEbiography}
\vspace{-1.2cm}

\begin{IEEEbiography}[{\includegraphics[width=1in,height=1.2in,clip,keepaspectratio]{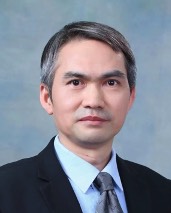}}]{Limin Xiao} 
received the B.S. degree in computer science from Tsinghua University, Beijing, China, in 1993, and the M.S. and Ph.D. degrees in computer science from the Institute of Computing, Chinese Academy of Sciences, Beijing, in 1996 and 1998, respectively.

He is currently a Professor with Beihang University, Beijing. He authored more than 260 papers on international journals and conferences. His research interests include computer architecture and system software, high performance computer and server system, system virtualization and cloud computing, big data storage and distributed file system and architecture and technology of intelligent computing chip.

Dr.Xiao is a CCF Distinguished Member.
\end{IEEEbiography}
\vspace{-1.2cm}

\begin{IEEEbiography}[{\includegraphics[width=1in,height=1.2in,clip,keepaspectratio]{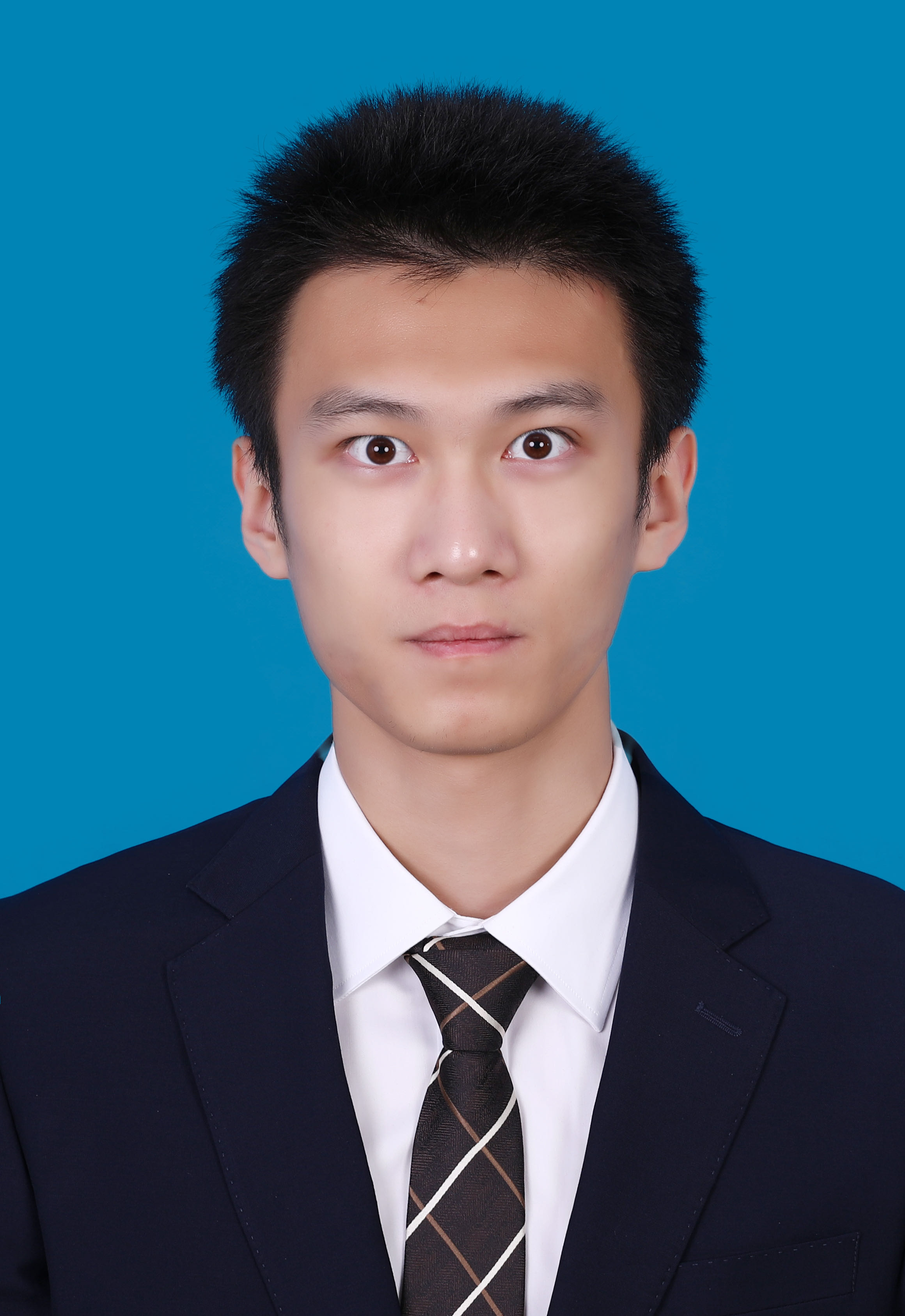}}]{Tianhao Cai} 
received the B.S. degree in Computer Science from the Beihang University, Beijing, China, in 2022. He is currently working toward the M.S. degree in Computer Science and Technology with the School of Computer Science and Engineering, Beihang University, Beijing, China. His research interests include computer architecture and deep learning accelerator.
\end{IEEEbiography}
\vspace{-1.2cm}

\begin{IEEEbiography}[{\includegraphics[width=1in,height=1.2in,clip,keepaspectratio]{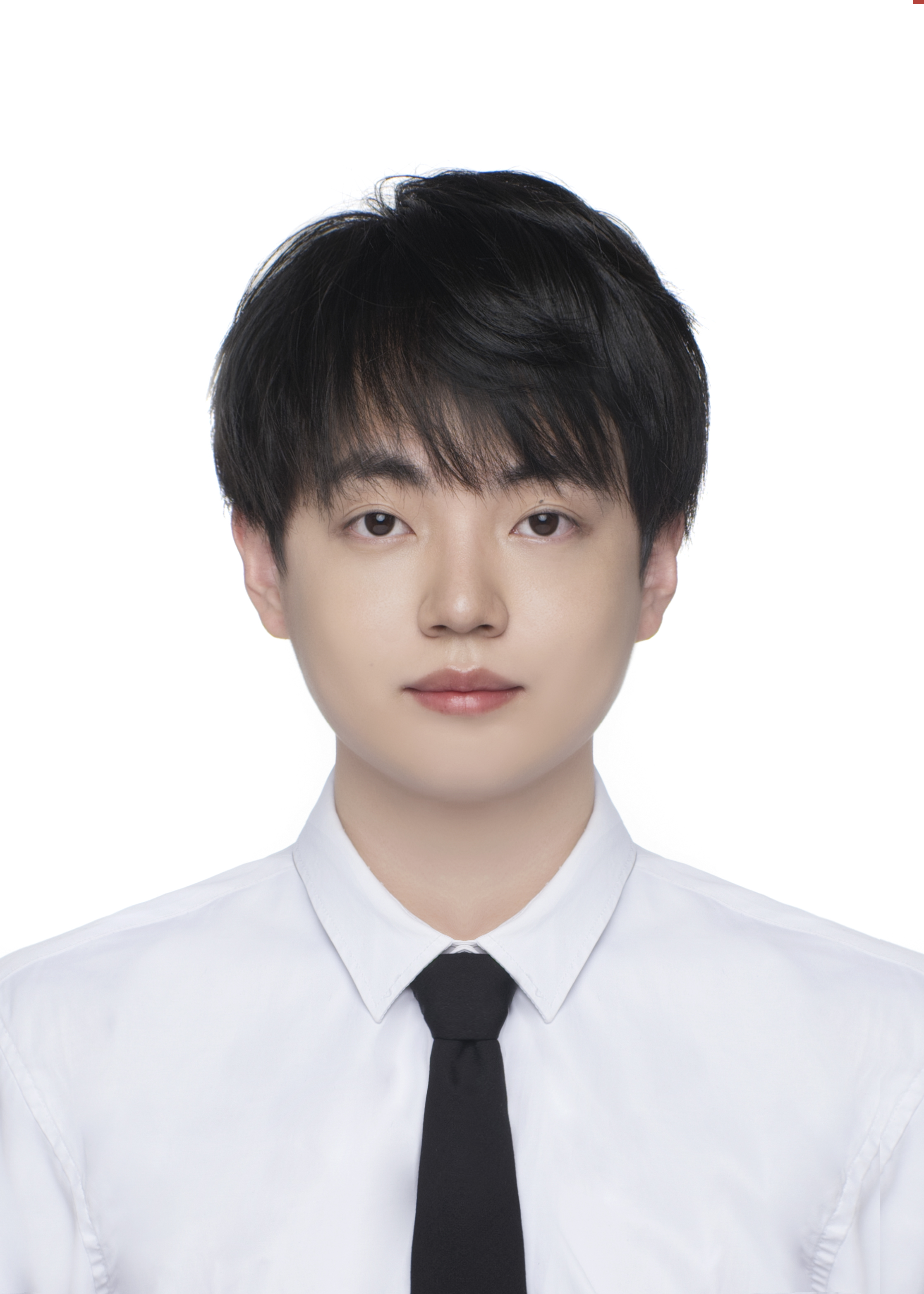}}]{Zeyu Wang}
received the B.S. degree in Computer Science and Technology from the Shandong University, Shandong, China, in 2022. He is currently working toward the M.S. degree in Computer Science and Technology with the School of Computer Science and Engineering, Beihang University, Beijing, China. His research interests include computer architecture and deep learning accelerator.
\end{IEEEbiography}
\vspace{-1.2cm}

\begin{IEEEbiography}[{\includegraphics[width=1in,height=1.2in,clip,keepaspectratio]{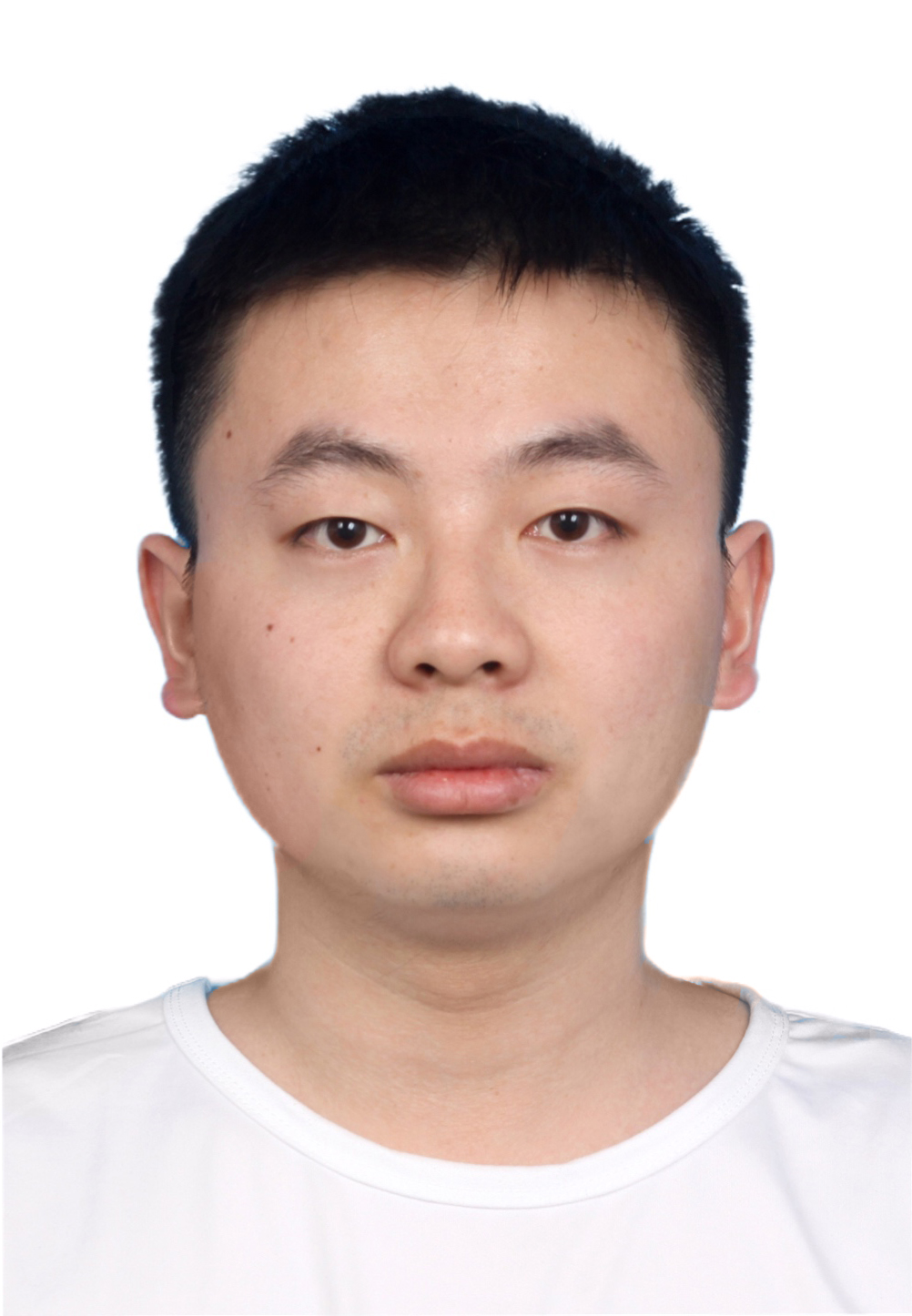}}]{Xiangrong Xu}
received the B.S. degree in Computer Science from the Beihang University, Beijing, China, in 2020. He is currently working toward the Ph.D. degree in Computer Architecture with the School of Computer Science and Engineering, Beihang University, Beijing, China. His research interests include computer architecture and GPU.
\end{IEEEbiography}
\vspace{-1.2cm}

\begin{IEEEbiography}[{\includegraphics[width=1in,height=1.2in,clip,keepaspectratio]{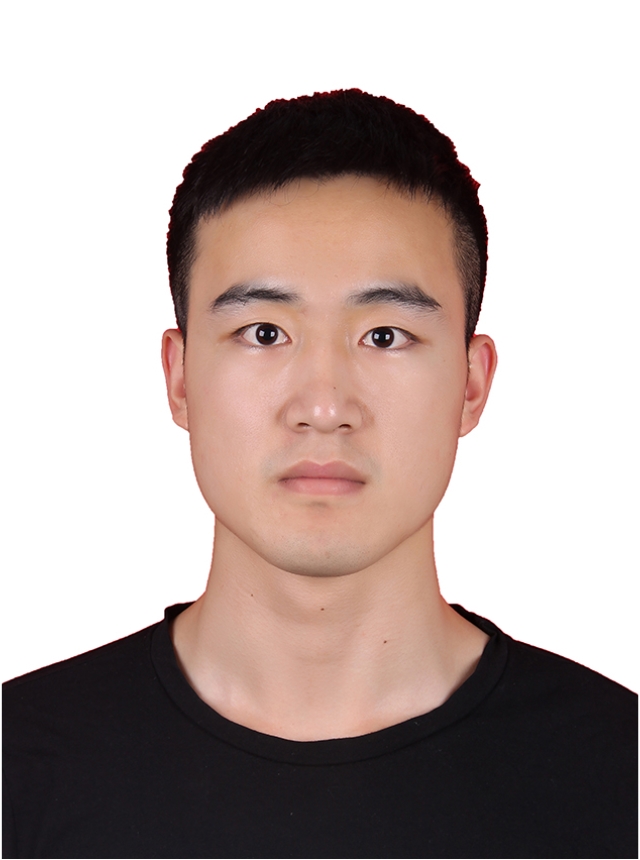}}]{Chenhao Zhang}
received the B.S. degree in Internet of Things Engineering from the China University of Petroleum (East China), Shandong, China, in 2019. He is currently working toward the Ph.D. degree in Computer Architecture with the School of Computer Science and Engineering, Beihang University, Beijing, China. His research interests include distributed file systems, storage system, high performance computing, replica technology.
\end{IEEEbiography}

\end{document}